\documentclass[review,3p]{elsarticle}

\usepackage{lineno}
\usepackage{hyperref}
\usepackage{amsmath, amssymb}
\usepackage{graphicx}
\usepackage{type1cm}
\usepackage{here}
\usepackage{color}
\usepackage{appendix}
\modulolinenumbers[5]

\journal{Physica A: Statistical Mechanics and its Applications}

\begin{document}

\begin{frontmatter}

\title{Estimation of crowd density applying wavelet transform and machine learning}

\author{Koki Nagao\corref{mycorrespondingauthor}}
\cortext[mycorrespondingauthor]{Corresponding author}
\address{Department of Aeronautics and Astronautics, School of Engineering, The University of Tokyo, 7-3-1, Hongo, Bunkyo-ku, Tokyo, 113-8656, Japan,}
\ead{k.nagao.0621@gmail.com}

\author[mymainaddress]{Daichi Yanagisawa}
\ead{tDaichi@mail.ecc.u-tokyo.ac.jp}

\author[mymainaddress]{Katsuhiro Nishinari}
\ead{tknishi@mail.ecc.u-tokyo.ac.jp}

\address[mymainaddress]{Research Center for Advanced Science and Technology, The University of Tokyo, 4-6-1, Komaba, Meguro-ku, Tokyo, 153-8904, Japan}

\begin{abstract}
We conducted a simple experiment in which one pedestrian passed through a crowded area and measured the body-rotational angular velocity with commercial tablets. Then, we developed a new method for predicting crowd density by applying the continuous wavelet transform and machine learning to the data obtained in the experiment. We found that the accuracy of prediction using angular velocity data was as high as that using raw velocity data. Therefore, we concluded that angular velocity has relationship with crowd density and we could estimate crowd density by angular velocity. Our research will contribute to management of safety and comfort of pedestrians by developing an easy way to measure crowd density.
\end{abstract}


\end{frontmatter}


  \section{Introduction}
Human crowd dynamics have been studied vigorously in terms of social and engineering science. To improve comfort or the safety in public area, it is important to estimate the state of pedestrian traffic quantitatively, i.e., density, velocity and travel time, and to predict pedestrian movement. However, due to the lack of such knowledge, many inefficient or dangerous situations may arise, e.g., stagnation of pedestrian flow in transportation systems in urban areas or evacuation of panicking crowds, leading to a stampede. 

The level of service (LOS) is frequently used as an index of the safety and comfort of pedestrians \cite{Fruin1971,Tanaboriboon1989,HCM2000,Asadi2013,Sahani2013,Muraleetharan2005}. Fruin defined the LOS based on the width of walkways and the volume of pedestrians \cite{Fruin1971}, which is considered the original definition. In research by Tanaboriboon and Guyano \cite{Tanaboriboon1989}, the LOS for walkways in Bangkok is calculated by the area occupied by pedestrians, where the concepts of Fruin's LOS are used in this calculation. In this paper, some kinds of LOS are developed to reflect a difference in environments using previous studies as a reference \cite{Sahani2013,Muraleetharan2005}. The Highway Capacity Manual (HCM)\cite{HCM2000} is often referenced as a guideline that considers the velocity of pedestrians, as well as the width of walkways and the volume of pedestrian traffic. This guideline presents some definitions of LOS in different environments, such as the case where pedestrians and bicyclists are mixed. As above, the density of pedestrians must be known to calculate most of the LOS. However, it seems difficult to apply the LOS which is complicated to calculate practically because there are some problems measuring feature values such as crowd density \cite{Asadi2013}. 

Video cameras are commonly used to record pedestrian movement over the long term. We can obtain much information about pedestrians, such as their positions, velocities and crowd densities from video data. However there are some problems with measuring crowd density using video cameras owning to their restricted setting. The area that can be recorded is usually limited by design of facilities or laws relating to the right to privacy. Moreover, some pedestrians may be concealed behind an obstacle or other pedestrians depending on the angle of the video camera. The limitation of recordable area and the impossibility of tracking pedestrians thoroughly are critical issues affecting accuracy. On the other hand, there are also problems of monetary and time costs. For example, when recording pedestrian movement in more than one place, we must prepare as many video cameras as there are places to be recorded, such that monetary and time cost may become high. Because solving the above issues allows the LOS to be calculated much faster, an easier method of extracting crowd density is more necessary than ever before.

Therefore, we regarded the crowd density as a feature value to be calculated from other feature values that can be easily measured. Recent studies have partially succeeded in modeling some collective pedestrian phenomena \cite{Yanagisawa2012,Gupta2015,Helbing2007,Townsend2014a,Chraibi2010,Lam2000,Zhang2014,Zhang2012a,Feliciani2016}, such that some relationships among feature values have been revealed. For example, a fundamental diagram is frequently considered to show the relationship among crowd density, velocity, and flow. The fundamental diagram, which can estimate pedestrian conditions, is used to compare crowd movements in different situations \cite{Helbing2007,Feliciani2016}. Moreover, the diagram is also used to estimate whether crowd movement in a simulation is valid compared to that in real situation \cite{Townsend2014a,Chraibi2010}. Thus, the fundamental diagram may be useful to estimate crowd density using velocity.

Then, we proposed a new method for estimating crowd density using the sensors in tablets \cite{Feliciani2016a}, most of which come with accelerometers and gyro sensors, as a simpler and less time-consuming method than using a video camera. By integrating the acceleration data from the accelerometer attached to pedestrians, we can obtain the velocities of pedestrians and calculate crowd density based on the fundamental diagram. However, accelerometers in tablets are less accurate than gyroscopic sensors. Therefore, we tried using only the gyro sensor to estimate crowd density by assuming a relationship with angular velocity. We obtained angular-velocity data from gyroscopic sensors on tablets attached to the torsos of the pedestrians. Then, the acquired data were preprocessed by a frequency analysis method using a continuous wavelet transform (CWT). Finally, we predicted the crowd density by applying machine learning to the preprocessed data. In this paper, we conducted real experiments to see whether our method is valid in a practical situation. In addition, we compared the performance of prediction by angular velocity data from tablet sensors and by velocity data from a video camera to confirm that the relationship between angular velocity and crowd density is as strong as that between velocity and crowd density.

The paper is structured as follows: We describe our experimental settings in section 2. We obtain angular velocity data of the torsos of participants in the experiments. Then, we introduce the wavelet transform, which is a time-series-analysis tools in section 3. In section 4, we explain methods of machine learning and forms of input data in order to deal with classification and regression. We discuss two definitions of crowd density, the global density, and the local density, which are addressed separately in the next two sections. First, we predict the global density, focusing on practicality rather than solution, in section 5. Next, we estimate the local density with a focus on gauging the degree of congestion around one pedestrian precisely in section 6. Finally, we conclude the study in section 7.


  \section{Experiment}


It is important to analyze the movement of crowds of various densities exhaustively when focusing on the relationship between this quantity and angular velocity. However, under certain circumstances, the density of crowds can change widely. Therefore, we conducted experiments in a simplified setup in order to control crowd density. 

\subsection{Setup}
\begin{figure}[h]
	\begin{center}
		\includegraphics[width=\linewidth]{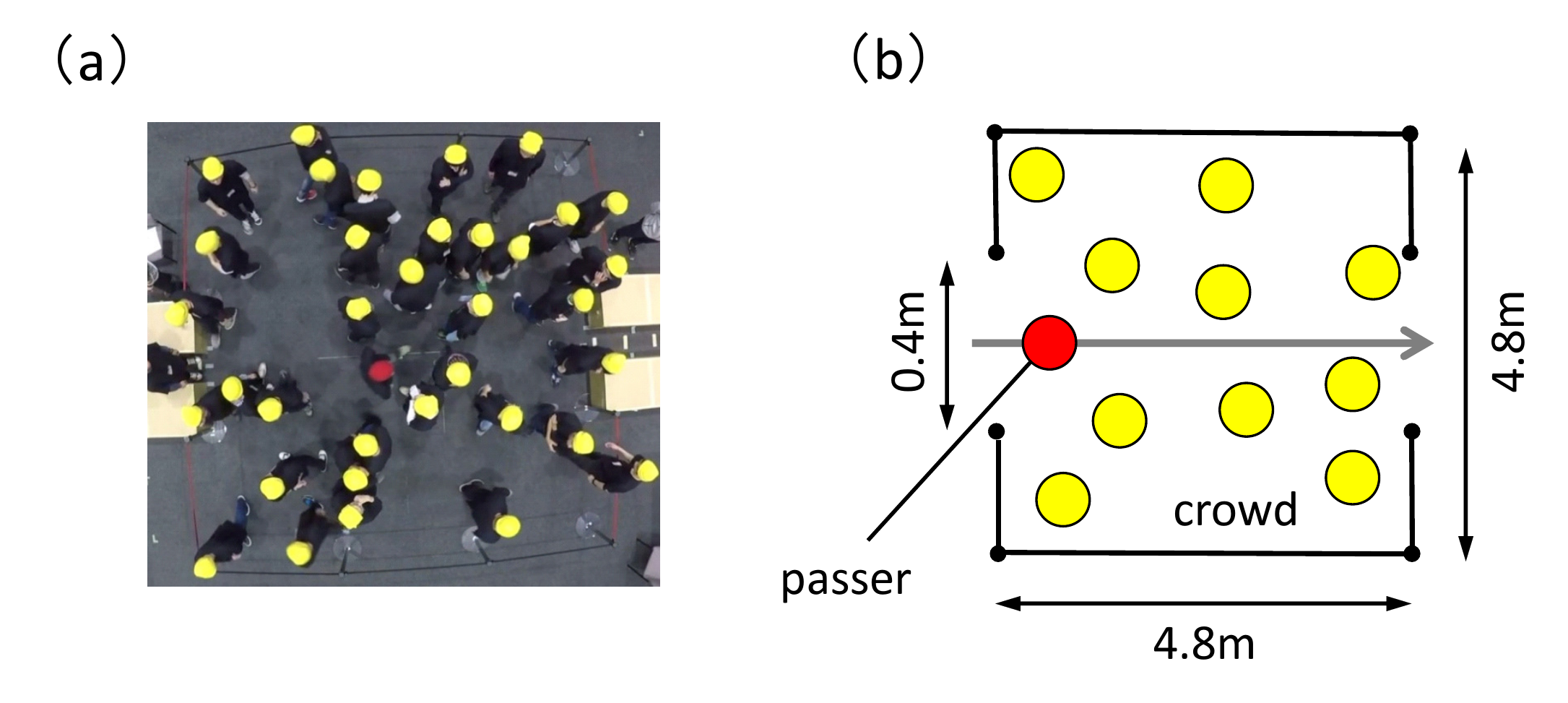}
	\end{center}
\caption{(a) Snapshot of the experiment. (b) Schematic diagram of the experiment.}
	\label{experiment_set}
\end{figure}


The experiments were conducted in a square area whose wall was composed of a belt partition  (Fig. \ref{experiment_set}). The length and width of the area were both 4.8 m, while the widths of the exit and the entrance were both 0.4 m. This area corresponds to a public open space such as a public square. Note that we chose an open space as our object of analysis because it is difficult to set a video camera there and tablet sensors may have an advantage.

Participants were divided into two groups, ``crowd members'' and ``passers''. The role of crowd members, who were issued yellow caps, is to walk randomly in the experimental area. The crowd members correspond to crowds in a public space. The number of crowd members was varied from 0 to 40 in increments of 5 (Table \ref{LOS}). Table \ref{LOS} shows that the range of densities of crowd members in this experiment covers almost the entire range of that in LOS \cite{Fruin1971}. We focus on the densities where pedestrians can move. Thus, \(\overline{\rho} = 1.78\ [m^{-2}]\), at which point the flow becomes unstable and potentially turbulent \cite{Fruin1971}, is high enough for our motivation. Moreover, according to the local density, we can observe a density higher than \(1.78\ [m^{-2}]\).

\begin{figure}[h]
	\begin{center}
		\includegraphics[width=\linewidth]{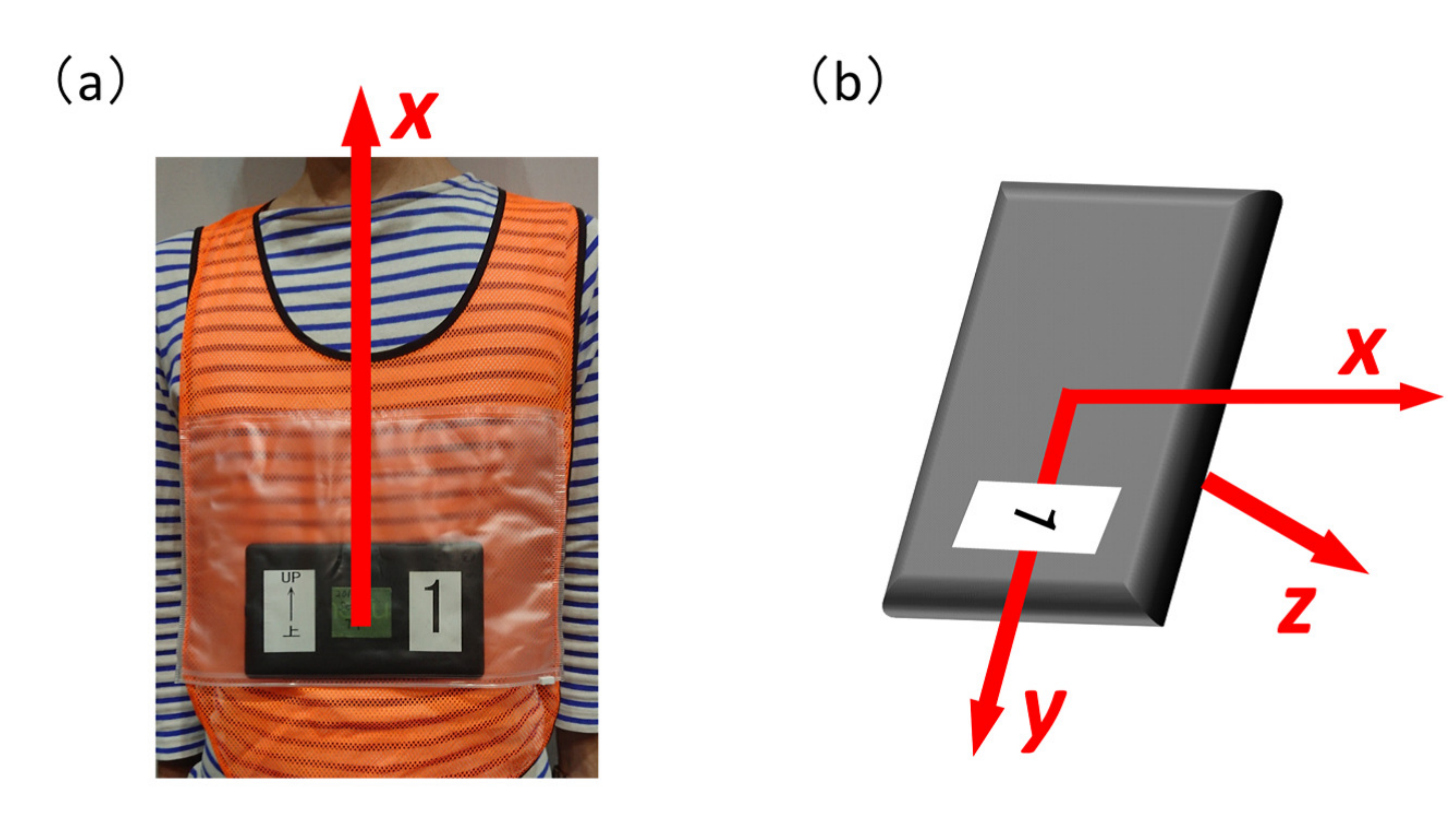}
	\end{center}
\caption{(a) Passer with a tablet. The passer wears a bib with a transparent pocket in which the tablet is placed. (b) Coordinates of the gyroscopic sensor in a tablet. Using the gyroscopic data with respect to the \(x\)-axis, we can track the rotation of the passer's torso.}
	\label{tablet}
\end{figure}

We instructed a passer, who wore a red cap, to walk from the entrance to the exit while passing through the crowd. Each sequence of this is defined as one trial. We conducted 240 trials while changing the crowd densities. The passer represents a pedestrian who knows his/her destination and the movement of the passer is the object of this analysis. The passer was equipped with a tablet (Nexus 7 (2013)) to measure the angular velocity of his/her torso (Fig. \ref{tablet} (a)). Note that the area was wide enough for the passer to be influenced little by boundaries. A video camera was installed over the area to record the entire experimental process.


The participants were male students from 18 to 28 years old who wore black T-shirts, allowing the positions of their heads to be analyzed more easily. We conducted similar experiments in which different students participated in different days, calling these experiments exp 1 and exp 2. These experiments included 240 and 90 trials (Table \ref{LOS}), respectively. In addition, participants are shuffled every 2-8 trials in order to reproduce a variety of crowd-member patterns. 

Note that we conducted similar experiments twice in order to estimate the generalization performance of our method of predicting crowd density. If the trend of the results of exp 1 is similar to that of exp 2, our method is less influenced by individual variability.

\begin{table}[htb]
  \caption{Overview of the experiments. The labels in classification indicate whether the global density is lower or higher than 0.93, where a trial in which the global density is lower than 0.93 \([{\rm m}^{-2}]\) has a label \(L\) and the other trial has a label \(H\). The label is mentioned in Sec. 4.3.2.}
\vspace{0.3cm}
\centering
  \begin{tabular} {cccccc}
\hline \hline
     \shortstack{Number of\\crowd members}  & \shortstack{Global\\density \(\overline{\rho}\) \([{\rm m}^{-2}]\)}  & LOS & \shortstack{Label in\\classification} & \shortstack{Trials of \\exp 1} &  \shortstack{Trials of \\exp 2} \\ \hline \hline
    0 & 0.0434 & A & \(L\) & 20 & -\\
    5 & 0.260 & A & \(L\) & 20 & - \\
    10 & 0.477 & B & \(L\) & 20 & 30 \\
    15 & 0.694 & C & \(L\) & 20 & - \\
    20 & 0.911 & D & \(L\) & 20 & 30 \\
    25 & 1.13 & E & \(H\) & 20  & -\\
    30 & 1.35 & E & \(H\) & 40  & 30\\
    35 & 1.56 & E & \(H\) & 40  & -\\
    40 & 1.78 & E & \(H\) & 40  & -\\ 
\hline \hline
\label{LOS}
  \end{tabular}
\end{table}

\break
\subsection{Acquired data}

In this subsection, we explain data acquisition and feature-value extraction for this experiment. Note that we used only the data while the passer's body was totally within the experimental area. The number of datasets, \(N\), is calculated according to \(N=({\rm sampling\ rate}) \times ({\rm time\ of\ experiment})\). \(N \approx 120 - 300\), depending on the trials.

\begin{description}
\item[Velocity \(v_t\)]\mbox{}\\
A video camera was installed over the room to record the entire experimental process. In order to analyze the video data, the software PeTrack \cite{Boltes2013} was used. This software automatically extracted the trajectories of the heads of participants, \({\bf r}_t\), from the recorded experimental video. The sampling rate was 30 Hz. Note that these trajectory data are used only to calculate the velocity and local density.

Velocity, \(v_t\), is calculated as follows:
\begin{equation}
v_t = \frac{\| {\bf r}_{t+6 \delta t} - {\bf r}_{t-6 \delta t} \|}{12 \delta t},
\end{equation}
\noindent
where \(\delta t\) is the time interval between neighboring datasets, in this case given by \(\delta t = 1/30\ [\rm sec]\). This definition plays the role as a filter, smoothing the datasets.
\end{description}

\begin{description}
\item[Angular velocity \(\omega_t\)]\mbox{}\\
We obtained time-series data concerning the angular velocity of a passer's torso. The sampling rate is 50 Hz. By focusing on the yaw (\(x\)-axis) of the angular velocity data, we can track the rotation of the passer's torso when seen from above.
\end{description}

\begin{description}
\item[Crowd density \(\overline{\rho},\ \rho_t\)]\mbox{}\\
In our research, the global density \(\overline{\rho}\) and the local density \(\rho_t\) are considered. The global density \(\overline{\rho}\) is defined as follows:

\begin{equation}
   \overline{\rho}\ [{\rm m}^{-2}]  = \frac{({\rm The\ number\ of\ crowd\ members}) + 1}{({\rm Floor\ space\ of\ the\ experimental\ area})\ [{\rm m}^2]}
\label{gd}
\end{equation}
\noindent

This density is an average crowd density in the experimental area, which is a constant 
value over one trial（Table \ref{LOS}）. Note that the number of participants in the experimental area, which corresponds to the numerator of eq. (\ref{gd}), is the sum of the number of crowd members and the passer. From the viewpoint of applying our method to the general public, knowledge of whether the area is generally crowded or not would be an adequate for detailed datasets of every moment. Thus, there is merit to using global density.


In contrast, the local density, \(\rho_t\), is the crowd density around a passer, which is calculated at each time steps. While there are several definitions of local density, we used the density distribution \cite{Steffen2010} in this study. The local density includes the effect of temporary bias in crowd density, reflecting the passer's feeling about the degree of congestion. The local density is useful for considering the relationship between feelings about the degree of congestion and the feature values at a certain moment. 

When investigating the relationship between the angular velocity, which is recorded by the tablet sensor (50 Hz), and the local density whose sampling frequency is equal to that of the video camera (30 Hz), we reduced the sampling rate of the angular-velocity data to 30 Hz. The angular velocity at an arbitrary moment is calculated as a weighted average of the two nearby datasets of angular velocity, where the weighting factor coefficient is the reciprocal of the time interval (Fig. \ref{cal_v}). Fig. \ref{time series} plots the angular velocity and the local density in one trial.


\begin{figure}[h]
	\begin{center}
		\includegraphics[width=10cm]{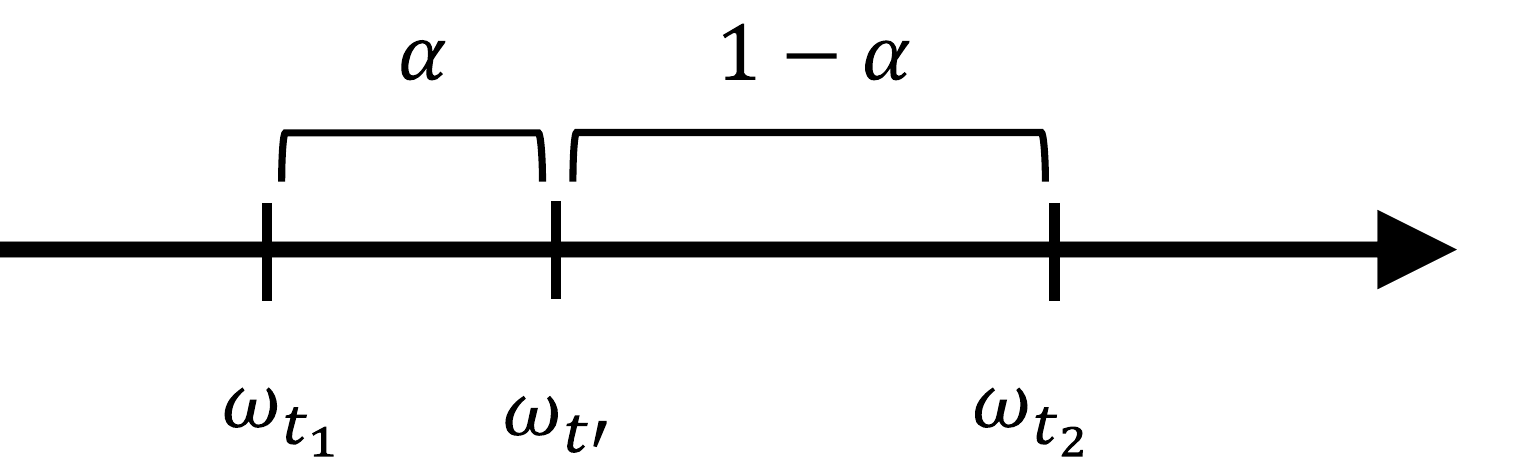}
	\end{center}
\caption{Conceptual diagram of down sampling of angular velocity. Now, the angular velocity at \(t'\) is calculated as a weighted average of the two near datasets of angular velocity, where the weighting factor coefficient is the reciprocal of the time interval. Thus, \(\omega_{t'}=(1-\alpha )\omega_{t_1} + \alpha \omega_{t_2}\), where the ratio of the time intervals is expressed as \(\alpha : 1-\alpha \).}
	\label{cal_v}
\end{figure}

\begin{figure}[h]
	\begin{center}
		\includegraphics[width=10cm]{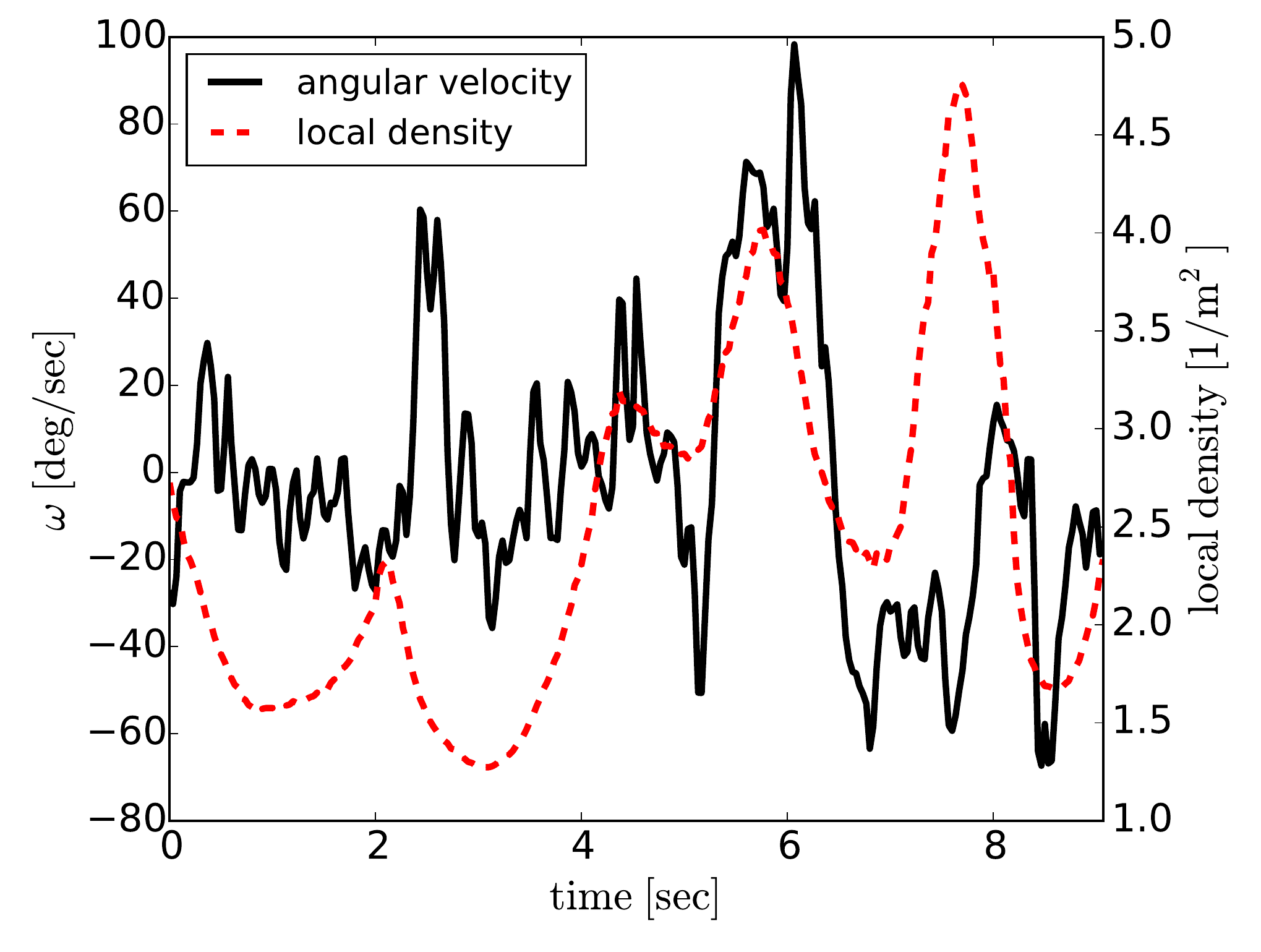}
	\end{center}
\caption{Experimental data from one trial in which there are 40 people in the crowd in exp 1. The black solid curve represents the angular velocity of a passer and the red dashed curve represents the local densities around the passer. Note that we regarded the density distribution as a local density, which we calculated using a Voronoi diagram \cite{Steffen2010}.}
	\label{time series}
\end{figure}
\end{description}


  \section{Time series analysis}
Time-series datasets are more strongly correlated with each other than the scattered datasets. In our experiments, an angular velocity cannot be dramatically different from its neighboring value, as we might intuitively expected. Therefore, we believe that it is more preferable to analyze based on correlation between neighboring datasets than analyze with datasets regarded as scattered datasets.

Frequency analysis, which is one method of considering time correlation, is often used to extract the frequency component from a signal. In our experiments, passers walked freely when there were a few pedestrians around them. This situation leads a cyclic rotation of the passers; thus, we decided that the frequency analysis is effective in our case. Moreover, passers sometimes rotated vigorously. For such a case, it is important to observe a large angular velocity is observed. Therefore, instead of Fourier transform, which is a of frequency analysis tool and does not include time-based information, we used the CWT, which considers both time and frequency. The advantage of the CWT is discussed in Appendix A.

In this section, we explain the CWT, an important frequency-analysis tool. Then, we applied the CWT to our experimental data. 

\subsection{Continuous Wavelet Transform (CWT)}
The CWT \cite{Torrence1995} is an analytical tool that incorporates the concepts of both time and frequency. We considered a time-series \(x_n\), with equal time spacing \(\delta t\) and localized time index \(n=1 \cdots N\). In addition, we used the wavelet function \(\psi_0(\eta)\), where \(\eta\) is a non-dimensional time parameter. Hence, the CWT coefficient, \(W_n\), of a discrete sequence \(x_n\) was defined as a convolution of \(x_{n}\) and the wavelet function \(\psi_0(\eta)\): 


\begin{equation}
W_n(s) = \sum^{N-1}_{n'=0} x_{n'} \left( \frac{\delta t}{s} \right)^{1/2} \psi_0 ^{\ast} \left( \frac{(n'-n) \delta t}{s} \right) ,
\label{CWT_def}
\end{equation}
\noindent
where \(s\) represents the scale, which is proportional to the reciprocal of the Fourier frequency and \(\ast\) indicates a complex conjugate. Now, the normalization of \(\psi_0\) is considered in this definition. The scale \(s\) can be any number greater than 0 and the localized index \(n\) is any natural number between 1 to \(N\). By varying the scale and the localized index, we can extract the degree of arbitrary frequency components at each time step. Note that the wavelet coefficient reflects the size of the values of \(x_{n}\). 

\subsection{Applying CWT to the experimental data}
We used the Mexican hat wavelet \cite{Torrence1995} as a wavelet function \(\psi_0(\eta)\):

\begin{equation}
\psi_0 (\eta) = \frac{2}{\sqrt{3\sigma} \pi^{1/4}} \left( 1 - \left( \frac{\eta}{\sigma} \right)^2 \right) e^{{-\frac{\eta^2}{2\sigma^2}}}
\label{mexh}
\end{equation}
\noindent
where \(\sigma\) is a parameter and we set \(\sigma=1\) in our study. 


In the previous subsection, we introduced the scale \(s\) as an equivalent value to the Fourier frequency. It is known that the relationship between the scale \(s\) and the Fourier frequency \(f\) can be analytically calculated with each wavelet function. In particular, the Fourier frequency of a Mexican hat wavelet is calculated as 

\begin{equation}
f = \sqrt{\frac{5}{2}} \times \frac{1}{2 \pi s} .
\label{fourier wavelength}
\end{equation}
\noindent
In the case of the Mexican hat wavelet with \(s=1\) and \(\sigma = 1\), the Fourier frequency is calculated as 0.252. Note that the frequency is \(0.252*30=7.56\) Hz when \(s=1\) and the sampling frequency is 30 Hz.

\begin{figure}[h]
	\begin{center}
		\includegraphics[width=\linewidth]{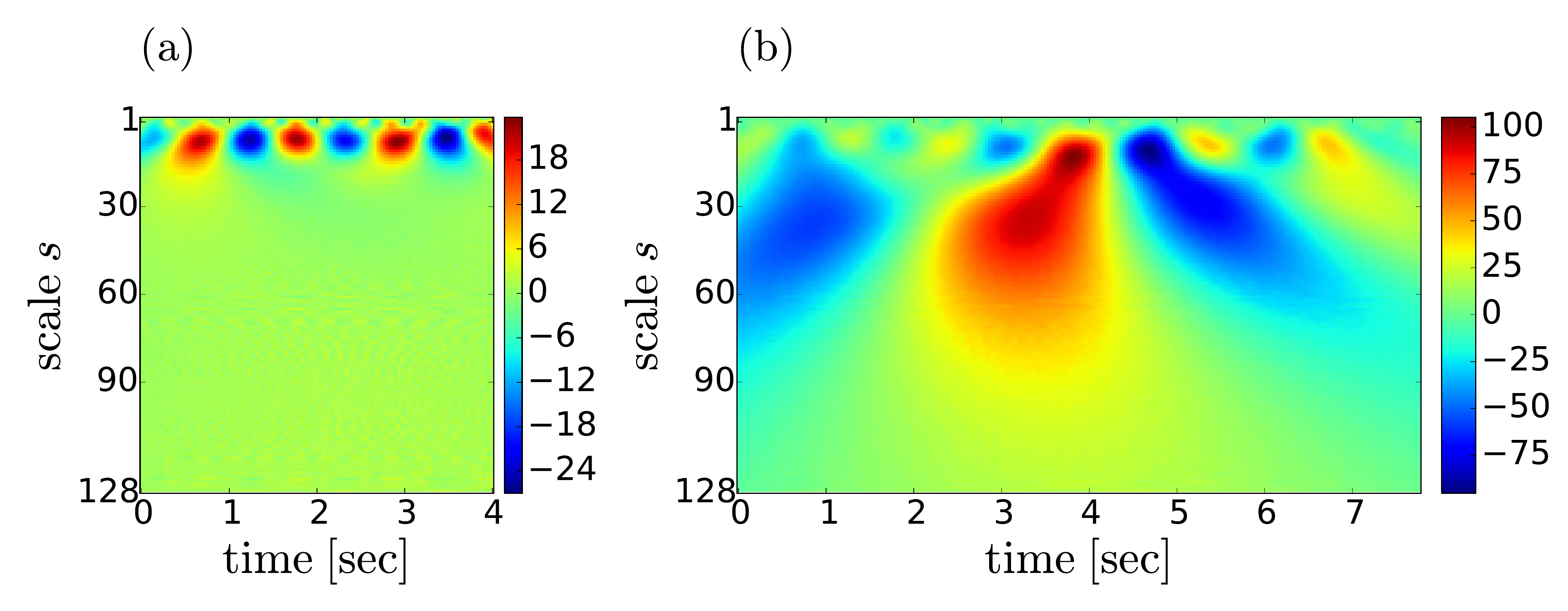}
	\end{center}
\caption{Scalogram of the gyroscopic data from one trial in a crowd of (a) 0 members and (b) 40 members. Note that the data from exp 1 were used. }
	\label{wavelet exp}
\end{figure}

In order to visualize the result of CWT, a scalogram is frequently drawn. This scalogram is a heat map whose \(x\)-axis represents time and whose y-axis represents the scale \(s\), and matrices are colored according to the wavelet coefficient \(W_n(s)\). Note that a larger scale leads to a lower frequency of the wavelet function.

We draw the scalogram by applying CWT to the experimental angular-frequency data \(\omega_n\ (n=0...N-1)\) (Fig. \ref{wavelet exp}). Note that the scale \(s\) varied within \( \{ s \mid 1 \leq s \leq 128 \ \land \ s \in \mathbb{Z} \} \). 

\begin{figure}[h]
	\begin{center}
		\includegraphics[width=10cm]{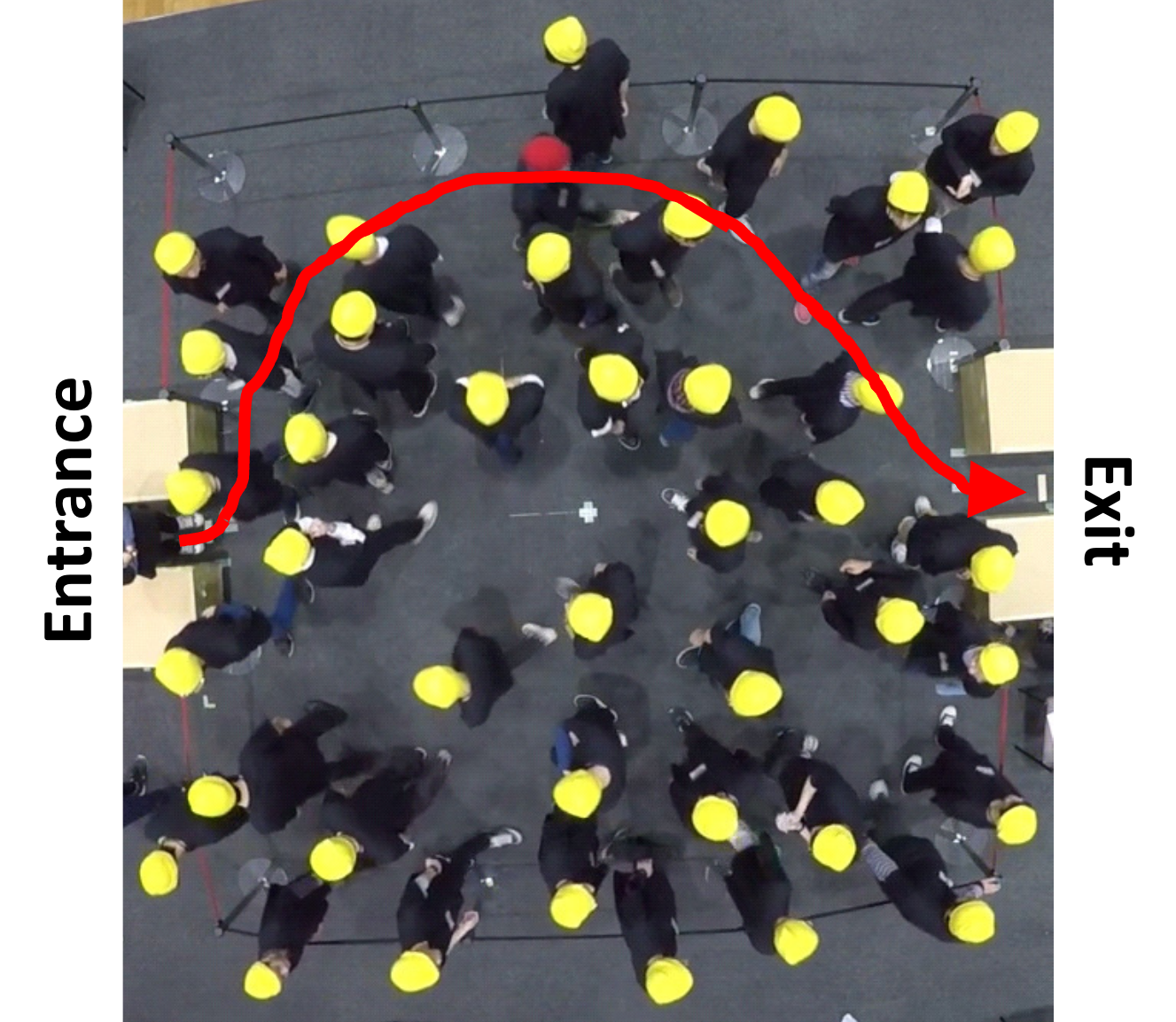}
	\end{center}
\caption{Snapshots of the case where the passer detours to avoid collisions with crowd members. The red curve is the schematic image of the passer's path. Note that the number of the crowd members is 40.}
	\label{detour}
\end{figure}

In both Figs. \ref{wavelet exp} (a) and (b), the cyclic fluctuation of the wavelet coefficient is observed along the line \(s = 10\) (the frequency is 0.755 Hz). This is because the passer walks freely with fluctuation of their torso, where the movement has a relatively high-frequency. Then, we focused on the low-frequency area in Fig. \ref{wavelet exp} (b). We can observe the fluctuation of the wavelet coefficient with \(s \geq 20\) (the frequency is less than 0.377 Hz). We considered the fluctuation pattern to reflect the movement which changed over a long time, due to such as processes as detouring in the experimental area without walking straight to the exit in order to avoid collisions with crowd members (Fig. \ref{detour}). Our hypothesis is supported by the fact that the fluctuation pattern cannot be observed in Fig. \ref{wavelet exp} (a), where the passer in the low-density area does not need go around.

  \section{Machine learning}
We used machine learning method in our study because it is suitable to handle comprehensively a wide variety of data, such as angular velocity and wavelet coefficient. We will now describe below how machine learning was employed in our study.

In a machine-learning algorithm, a dataset is a feature vector \(\bf{x}\) with a label \(y\), with a number of elements given by \(d\). A feature vector can be interpreted as one point in a \(d\)-dimensional feature space. Now, the purpose of machine learning is to predict the labels of feature vectors \({\bf x}_{\rm test}\) whose labels are unknown by using other datasets whose feature vectors \({\bf x}_{\rm train}\) and labels \(y_{\rm{train}}\) are known. The known and unknown data are called training and test data, respectively. Finally, we estimated the performance of the machine learning algorithm by comparing the predicted labels, \(y_{\rm{pred}}\), and the true labels, \(y_{\rm{test}}\). 

\subsection{\(k\)-nearest-neighbor ($k$-NN) algorithm}
The \(k\)-nearest-neighbor ($k$-NN) algorithm is an example of a machine-learning algorithm \cite{Dudani1976,Devroye1994}. It can be used for both classification and regression. Note that a label has a discrete value or a class name in classification, and a label has a continuous value upon regression. The \(k\)-NN algorithm is executed as the following steps:

\begin{enumerate}
 \item Place feature vectors of training data \({\bf x}_{\rm train}\) and one feature vector of test data \({\bf x}_{\rm test}\) in the feature space.
 \item Extract the \(k\) nearest training datasets from the \({\bf x}_{\rm test}\). Note that \(k\) can be any natural number less than the number of training datasets, but is generally an odd number.
 \item In classification, the most common label of the \(k\) nearest training datasets is regarded as the predicted label \(y_{\rm{pred}}\). In regression, the mean value of the labels of the \(k\) nearest training datasets is regarded as the predicted label \(y_{\rm{pred}}\). Note that weighting was introduced to the algorithm in the most common definition. In classification, we sum up the inverse distances between the feature vector of the test dataset and the training datasets for each label, and the label with the largest sum is the predicted label \(y_{\rm{pred}}\). In regression, the average value of the labels of \(k\) nearest training datasets weighted by the distance between the feature vector of the test and training datasets is regarded as \(y_{\rm{pred}}\).
\item Repeat steps 1\(\sim\)3 for all test datasets.
\end{enumerate}

\subsection{Cross-validation}

\begin{figure}[h]
	\begin{center}
		\includegraphics[width=10cm]{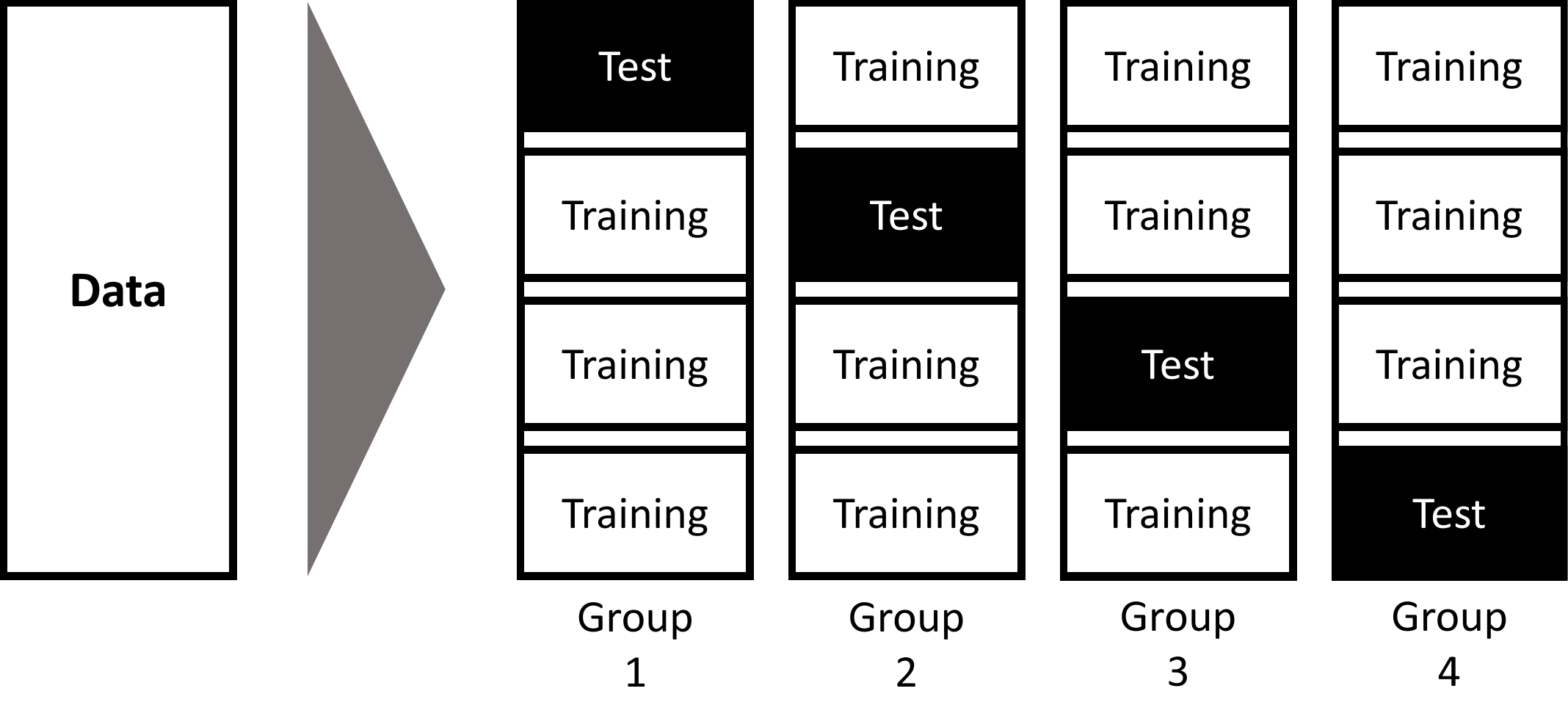}
	\end{center}
\caption{Conceptual scheme of cross-validation of classification in the case where \(m=4\). For example, when prediction accuracy is 75\%, 65\%, 95\% and 85\%, respectively, the accuracy of the machine algorithm is calculated as \(80\%\).}
	\label{cross_validation}
\end{figure}

Cross-validation is a method used to estimate the performance of a machine-learning algorithm using obtained data (Fig. \ref{cross_validation}). First, we divide the obtained datasets into \(m\) groups. One group is considered to be a test dataset and the other groups are considered training datasets. Then, we predict the labels of the test datasets using machine learning as discussed in the previous subsection. We repeat these processes while changing the group of test data by finishing all combinations. During classification, we calculate the mean value of the success probability of \(m\) predictions as the performance of the machine algorithm. Note that \(m\) can have any natural number less than the number of datasets obtained. The larger \(m\) is, the more accurately we can estimate the performance of the machine-learning algorithm; however, the calculation time becomes longer because the repetition time increases.

\subsection{Style of experimental data}
\subsubsection{Feature vector}
In this subsection, we explain the data style in density prediction. Firstly, we introduce the following feature vectors \({\bf x}\), which are the focus of this study (Table. \ref{feature}). 

\begin{description}
\item[raw velocity \(v_n\)]\mbox{}\\
Because the relationship between velocity and crowd density has been confirmed through such means as fundamental diagrams, we presumed that the crowd density could be predicted by velocity data. One of our purpose is to compare the results of prediction by velocity and by angular velocity based on the presumption. The data style of the raw velocity is considered to be \({\bf x}_1=[v_n]\), the dimension \(d\) of the datasets is 1, and the number of datasets is \(N-12\) per trial.
\end{description}

\begin{description}
\item[raw angular velocity \(\omega_n\)]\mbox{}\\
In Fig. \ref{time series}, the correlation between the angular velocity and the local density can be observed; thus, we presumed that the crowd density can be predicted by raw angular-velocity data. The style of the raw angular-velocity data is considered to be \({\bf x}_2=[\omega_n]\), the dimension \(d\) of datasets is 1 and the number of the datasets is \(N\) per trial.
\end{description}

\begin{description}
\item[maximum angular velocity, \(\omega_{\rm max}\)]\mbox{}\\
We focused upon the sharp rotation of the torso of the passer as a characteristic movement when passing through highly dense crowds in order to avoid collisions with others. Sharp rotation, which decreases the effective radius of the passer, cannot be observed under free-flow conditions; thus, it may be considered a sign of a high-density condition. In order to detect the timing of sharp rotation, we focused on the large-angular-velocity condition because we presumed that sharp rotation may lead to larger angular velocities than those in the free-flow condition. Thus, we regarded the maximum angular velocity in a trial as a feature value of the passer:
\begin{equation}
\omega_{\rm{max}} = {\rm max}(|\omega_n|)
\end{equation}
\noindent
The data style of the maximum angular velocity is considered to be \({\bf x}_3=[\omega_{\rm{max}}]\), the dimension \(d\) of the datasets is 1, and the number of datasets is \(1\) per trial.
\end{description}

\begin{description}
\item[wavelet coefficient, \({\bf W}_n\)]\mbox{}\\

We extracted frequency and time-based information from the angular-velocity time-series data using the CWT. The CWT can detect peaks in the time-series data, such that strong rotation of the passer's torso is reflected in the analysis. Thus, we regarded the wavelet coefficient as a feature value and the data style to be shown in the following equation

\begin{equation}
\begin{split}
 {\bf W}_n  &= \rm\it [W_{n-l}({\rm 1}),\ W_{n-l}({\rm 2}),\ \cdots,\ W_{n-l}({\rm 128}),\ \cdots,\ W_{n}({\rm 1}),\ \cdots,\ W_{n}({\rm 128}),\\
&\quad \ \cdots,\ W_{n+l}({\rm 1}),\ \cdots,\ W_{n+l}({\rm 127}),\ W_{n+l}({\rm 128})],
\end{split}
\end{equation}
\noindent
where this feature vector includes datasets during \(2l+1\) time steps. Therefore, we concluded that the cyclic fluctuation of the wavelet coefficient along the time axis can be reflected in the machine learning algorithm. 
\begin{figure}[h]
	\begin{center}
		\includegraphics[width=\linewidth]{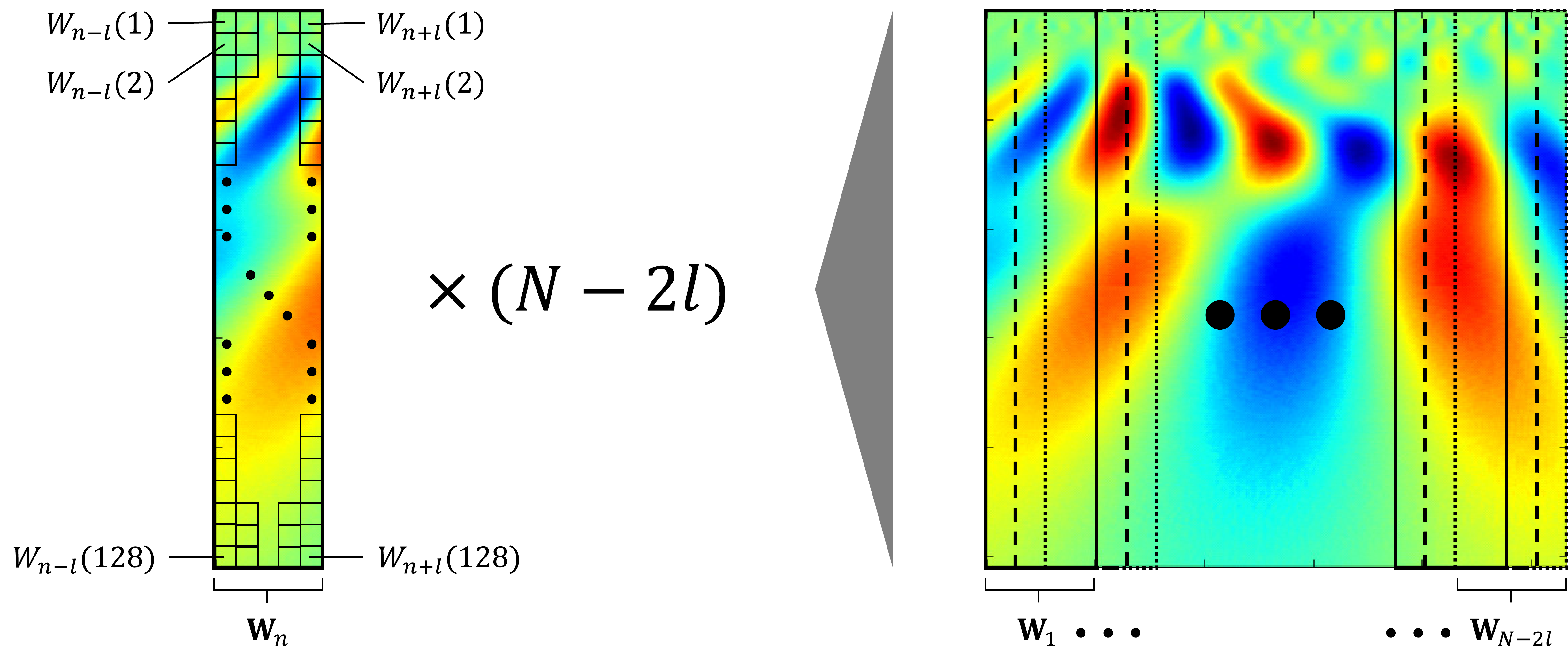}
	\end{center}
     \caption{Data style of the wavelet coefficient \({\bf W}_n\). \({\bf W}_n\) is a 1-D vector containing \(128 \times (2l+1)\) elements.}
	\label{datastyle_wavelet}
\end{figure}

Note that we set \(l=7\) in this study. Moreover, we regarded \({\bf W}_n\) as the feature vector \({\bf x}_4\), where the dimension \(d\) of datasets is \(2l+1\) and the number of \({\bf x}_4\) per trial is \(N-2l\) (see Fig.\ref{datastyle_wavelet}). 
\end{description}

\begin{description}
\item[sequential velocity, \({\bf V}_n\)]\mbox{}\\
The wavelet coefficient \({\bf x}_4\) contains information concerning certain time steps, allowing \({\bf x}_4\) to reflect the time correlation of the angular-velocity data in the machine-learning algorithm. Also, in the case of predictions made using velocity, we considered the feature vector containing information about these time steps, as shown in the following equation;

\begin{equation}
 {\bf V}_n = \rm\it [v_{n-l},\ v_{n-l+{\rm1}}, \cdots ,\ v_{n+l}].
\end{equation}
\noindent
In analogy with the case of the wavelet coefficient \({\bf x}_4\), cyclic fluctuation of the velocity along the time axis can be reflected in the machine-learning algorithm. Note that we set \(l=7\) in this study. Moreover, we regarded \({\bf V}_n\) as the feature vector \({\bf x}_5\), where the dimension \(d\) of the datasets is \(2l+1\) and the number of \({\bf x}_5\) values per trial is \(N-12-2l\).
\end{description}

\begin{description}
\item[non-sequential velocity, \({\bf V}_n'\)]\mbox{}\\
We prepared the sequential velocity \({\bf x}_5\) because the feature vector, which contains the sequential time series information, reflects the trend of the raw velocity data precisely. As a target for comparison, we considered the non-sequential velocity \({\bf V}_n'\) as the feature vector \({\bf x}_6\), which contains data from certain time steps with the steps chosen randomly. By comparing the results of \({\bf x}_5\) and \({\bf x}_6\), we estimated the influence of autocorrelation. The dimension and number of \({\bf x}_6\) values per trial are set to be the same as the number of \({\bf x}_5\) values per trial.
\end{description}

\begin{table}[htb]
 \centering
 \caption{Summary of the feature values}
 \begin{tabular} {cccccc}
 \({\bf x}_i\) & Feature value & Feature vector & \(d\) & Num of datasets \\ \hline \hline
 \({\bf x}_1\) & raw velocity & \([v_t]\) & 1 & \(N-12\) \\
   \({\bf x}_2\) & raw angular velocity & \([\omega_t]\) & 1 & \(N\) \\
    \({\bf x}_3\) & maximum angular velocity & \([\omega_{\rm{max}}]\) & 1 & 1 \\
   \({\bf x}_4\) & wavelet coefficient & \({\bf W}_n\) & \(2l+1\) & \(N-2l\) \\
   \({\bf x}_5\) & sequential velocity & \({\bf V}_n\) & \(2l+1\) & \(N-12-2l\) \\
   \({\bf x}_6\) & non-sequential velocity & \({\bf V}_n'\) & \(2l+1\) & \(N-12-2l\) \\
\label{feature}
 \end{tabular}
\end{table}

\subsubsection{Label}
Next, we explain the labels, which are the target of our prediction. We considered two types of labels according to the characteristics of the global and the local densities.

\begin{description}
\item[global density \(\overline{\rho}\)]\mbox{}\\
In predicting the global density, the labels \(y\) were constant during each trial because the global density is constant during each trial. We treated the prediction of the global density as a classification problem. The datasets of the trials in which the global density is 0.93 \([{\rm m}^{-2}]\) or more are labeled ``high-density'' (\(H\)) and the other datasets have the low-density label (\(L\)). Note that we decided the threshold based on the LOS, where \(\rm LOS=E\) (the global density being larger than 0.93 \([{\rm m}^{-2}]\)) is the condition for which transition to the turbulent phase may occur and 0.93 \([{\rm m}^{-2}]\) is suitable for such a threshold (Table \ref{LOS}). We consider the number of labels to be sufficient because it may be the most important factor in detecting the transition from a low-density condition to high-density condition in a practical way. For example, in a dense event space, it is preferable for a manager to guide pedestrians to a less-crowded space when the crowd density exceeds a certain threshold value. This is because a guidance is not necessary when the space is not crowded, and a speedy guidance is necessary when the crowd density becomes too high to assure pedestrians' comfort. Therefore, we consider this to be necessary and sufficient to distinguish the high-density condition from the low-density condition.
\end{description}

\begin{description}
\item[local density \(\rho_n\)]\mbox{}\\
On the other hand, in the case of local-density prediction, we used a different labeling method because the local density is calculated at each time step. We treated the prediction of the local density as a regression problem. Using this density, we can investigate the relationship between the rotation of a passer's torso and the degree of congestion felt by the passer in detail. In the case of datasets that contain information about a single time step, the label \(y\) is set to be the local density \(\rho_n\) at the present moment. In the case of the datasets containing information about \(2l+1\) time steps \([n-l,\ n+l]\), the label is set to be the weighted mean of the local density during the corresponding steps. We used the normal distribution as a weighting function. The label \(y\) when the localized index is \(n\) is defined as following equation;

\begin{equation}
y = \frac{  \sum_{n'=n-l}^{n+l} \frac{1}{\sqrt{2 \pi}} \exp\left(-\frac{(n'-n)^2}{2}\right) \times \rho_{n'}  }
{  \sum_{n'=n-l}^{n+l} \frac{1}{\sqrt{2 \pi}} \exp\left(-\frac{(n'-n)^2}{2}\right)  }
\end{equation}

\end{description}

  \section{Global density}
\subsection{Classification}
By using the feature vectors \({\bf x}_i\ (i=1,\ 2,\ 3,\ 4,\ 5,\ 6)\) in the previous section, we developed an algorithm to estimate the global densities in the trials.

In order to conduct cross-validation, we divided all the data from exp 1 into 240 groups (\(m=240\)), the same as the number of trials of exp 1. Data in 239 groups are assigned as training data and data in the other group are assigned as test data. This kind of the cross-validation is called leave-one-out cross-validation.

We used a \(k\)-NN algorithm to predict the labels of test data, with \(k=51\). Note that the above-mentioned weighting was introduced to the algorithm.

We allocated a label \(L\) or \(H\) to the test trials as a result of prediction. If there are 2 or more data points in one trial, we applied the \(k\)-NN algorithm to all data points and the label of that trial is selected as the mode of the prediction results. Then we decided whether the prediction was a success or failure by comparing the true and predicted labels. 

The prediction of each trial is based on the movement of one passer. However, in practical cases, more than one pedestrian will walk through the area in which we measure crowd densities. Therefore, we developed a density-estimation method capable of considering feature values of \(N_{\rm passer}\) pedestrians comprehensively. This method works as follows: firstly, we chose \(N_{\rm passer}\) trials in which the number of crowd members is the same. Then, we regarded the mode of the predicted labels of the trials as the total prediction result of the \(N_{\rm passer}\) trials. 

This method is called ensemble training. If the estimation accuracy of one trial is more than \(50\%\), it is known that a larger \(N_{\rm passer}\) leads to higher accuracy according to Condorcet's jury theorem \cite{jury}. Note that we conducted analysis with \(N_{\rm passer}=5\).

In addition, we conducted another validation in which the exp-1 data were regarded as training data and exp 2 was regarded as test data. This validation is based on an assumption that the relationship between crowd density and the movement of passers in exp 1 is similar to that of exp 2. By comparing results of this validation and the previously mentioned cross-validation, we estimated the influences of individual characteristics upon our method's accuracy.

\subsection{Confusion matrix}
The confusion matrix, a table showing the number of true and false predictions, is frequently used to evaluate classification-algorithm performance. In the case of 2-class classification, we can summarize this matrix as Table \ref{confusion_matrix}. 

\begin{table}[htb]
  \centering
  \caption{Confusion matrix}
  \begin{tabular} {c|c||c|c|}
   \multicolumn{2}{c||}{} & \multicolumn{2}{|c|}{Predicted label} \\ \cline{2-4}
     & & A & B \\ \hline \hline
    True label & A & True A& False B \\ \cline{2-4}
     & B & False A& True B \\ \hline
  \end{tabular}
\label{confusion_matrix}
\end{table}

From the confusion matrix, we can calculate the estimation indices of classification performance using the following equations:

\begin{eqnarray}
{\rm Accuracy} &=&{\rm \frac{{\it N}_{TrueA} + {\it N}_{TrueB}}{{\it N}_{TrueA} + {\it N}_{TrueB} + {\it N}_{False} + {\it N}_{FalseB}}},\\
\rm Sensitivity\ of\ A &=&{\rm \frac{{\it N}_{TrueA}}{{\it N}_{TrueA} + {\it N}_{FalseB}}},\\
\rm Precision\ of\ A &=&{\rm \frac{{\it N}_{TrueA}}{{\it N}_{TrueA} + {\it N}_{FalseA}}}
\end{eqnarray}
\noindent
where \(N_x\) is the number of samples corresponding to \(x\).
The accuracy is defined as the ratio of correct predictions, which is calculated using all data without distinguishing among labels. The sensitivity of label A is defined as the proportion of label-A-prediction cases that are accurate. This represents the probability of detecting the label A without misses. On the other hand, the precision of label A is defined as the proportion of true label-A cases that were predicted. The larger this value, the more reliable the prediction results. Note that the sensitivity and precision of label B are also calculated in the same manner.

\subsection{Results and Discussion}
\subsubsection{Performance estimation: data for exp 1}
\begin{table}[htb]
  \centering
  \caption{Cross-validation results using exp 1}
\scalebox{0.9}{
\begin{tabular}{p{0.01\linewidth}p{0.29\linewidth}p{0.11\linewidth}p{0.11\linewidth}p{0.11\linewidth}p{0.11\linewidth}p{0.11\linewidth}}
 \({\bf x}_i\) &  Feature value& Accuracy [\%]  & Sensitivity of \(L\) [\%] & Sensitivity of \(H\) [\%]  & Precision of \(L\) [\%] & Precision of \(H\) [\%] \\ \hline \hline
 \({\bf x}_1\) &   raw velocity & 83.3 & 60.0 & 100.0 & 100.0 & 77.8 \\ 
 \({\bf x}_2\) &   raw angular velocity & 58.3 & 0 & 100.0 & - & 58.3 \\ 
 \({\bf x}_3\) &   maximum angular velocity & 83.3 & 60.0 & 100.0 & 100.0 & 77.8 \\ 
 \({\bf x}_4\) &   wavelet coefficient & 87.5 & 70.0 & 100.0 & 100.0 & 82.4 \\
 \({\bf x}_5\) &   sequential velocity & 87.5 & 70.0 & 100.0 & 100.0 & 82.4  \\ 
 \({\bf x}_6\) &   non-sequential velocity & 89.6 & 80.0 & 96.4 & 94.1 & 87.1 \\
\end{tabular}
\label{accuracy_vali}
}
\end{table}

Table \ref{accuracy_vali} shows the cross-validation results for exp 1. The sensitivity of label \(H\) is \(100\%\) in the case of all feature values except the non-sequential velocity. This means that we can invariably detect high-density situations. Some applications have been conceived using the strength of a feature value. For example, safety-management systems issue alerts when the density becomes high and are helpful for guiding pedestrians at a suitable moment. However, if the system frequently issues alerts in low-density situations by mistake, the algorithm is not practical for the real world. Therefore, we focused on the precision of label \(H\). A higher precision causes the frequency of false alarms to decrease. In predictions based on raw- and sequential-velocity data, the precision of \(H\) is relatively high. This result is natural because some relationships between velocity and density have already been revealed. 

On the other hand, the precision of \(H\) in the case using wavelet coefficients is as high as that using velocity. This fact suggests that extracting frequency components by CWT is effective in our method as hypothesized in Sec. 4. Although the precision of \(H\) using raw angular-velocity data is lower than that of the other cases, feature values properly preprocessed from angular velocity data are effective in our estimation method. Conversely, our result suggests a relationship between angular velocity and crowd density.

It is worth noting that the prediction based on maximum angular velocity offers high performance, although the amount of data is less than that for other feature values. As previously noted, we formed a hypothesis that the characteristic movement of pedestrians passing through crowded areas is sharp rotation which decreases the effective radius of the passer. This result proves the validity of our hypothesis. 

\begin{figure}[h]
	\begin{center}
		\includegraphics[width=10cm]{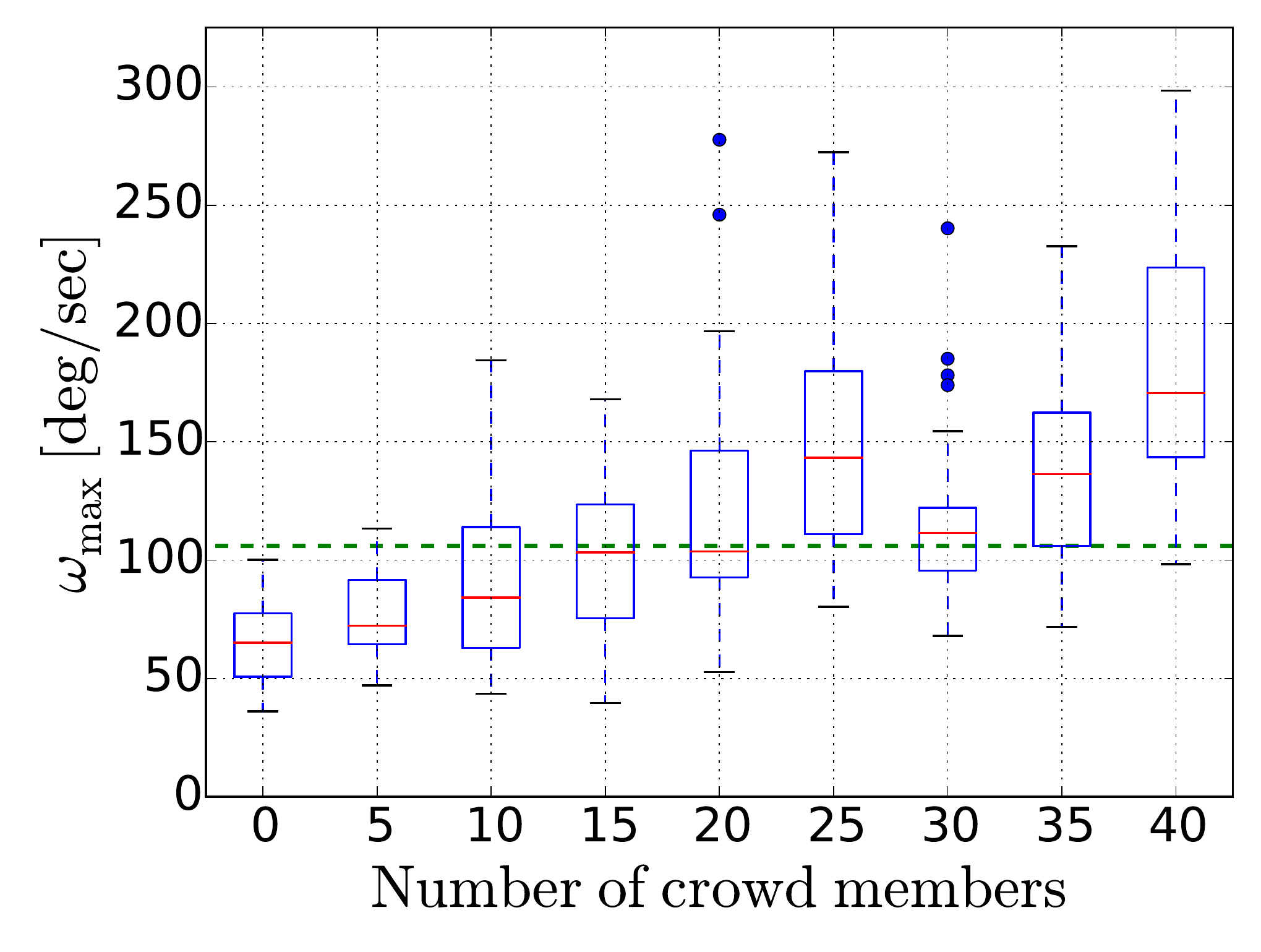}
	\end{center}
\caption{Box-and-whisker plot for \(\omega_{\rm max}\) in each trial. The bottom and top of the box are first and third quartiles, respectively, and the red band inside the box is the median. The ends of the whiskers represent the lowest and highest data points still within 1.5 IQR of the lower and upper quartiles, respectively. The outliers are plotted as round points. The green dashed line represents the threshold dividing high-density and low-density situations, where trials in which the maximum angular velocity exceeds the threshold are regarded as having high density. Note that we decided the threshold in such a way that the prediction accuracy was the highest.}
	\label{peak threshold}
\end{figure}

We also examined the maximum angular velocity \({\bf x}_3=[\omega_{\rm max}]\) in another way. Fig. \ref{peak threshold} shows box-and-whisker plots of \(\omega_{\rm max}\) in each trial, where the horizontal axis represents the number of crowd members. \(\omega_{\rm max}\) in trials in which crowd has 25 or more members is slightly larger than that in the other trials. Therefore, we developed a classification method in which we regard a trial as having high density if \(\omega_{\rm max}\) exceeds a threshold. The green dashed line in Fig. \ref{peak threshold} represents the threshold, which is \(106\ {\rm deg/sec}\). Note that we decided the threshold in such a way that the prediction accuracy was highest. Given the fact that the sensitivity of label \(H\) was \(89.3\%\), we believe the method in which only \(\omega_{\rm max}\) is used to be practical with a certain degree of accuracy. 

Next, we considered the sequential velocity. Firstly, by comparing sequential- and raw-velocity results, we find that feature vectors that contain information concerning certain time steps are better suited for the density-estimation method. Then, we compared results using the sequential and non-sequential velocities. The sensitivity of label \(H\) is higher using the sequential velocity data than using the non-sequential velocity data. One factor affecting in the difference between sequential and non-sequential velocity is that the dispersion of elements of the non-sequential velocity is larger than that of the sequential velocity. The sequential velocity has temporary bias in that all elements in a feature vector are large due to sharp rotation of the passer. On the other hand, the non-sequential velocity is composed of data for separate time steps, such that feature vectors are less biased than sequential velocity. This smaller bias is interpreted as there being less difference between each non-sequential-velocity vector, and this factor makes it difficult to detect sharp rotation of the passer, which is sign of high-density situations. It is difficult to decide which feature vector is better; however, considering only cases where detection of high density is emphasized for safety management, we can say that prediction by sequential velocity is more effective.

The prediction performance using wavelet coefficients is equal to or better than that using other feature values. On the other hand, prediction using the maximum angular velocity has an advantage in that this feature vector can be easily extracted. Therefore, we should choose which feature value is used according to each situation. Note that the above results were not greatly changed with the other parameters, which are \(k\) in the \(k\)-NN algorithm and \(l\) of \({\bf x}_4\), \({\bf x}_5\), \({\bf x}_6\). This is discussed in detail in Appendix B.

\subsubsection{Performance estimation: data for exps 1 and 2}
We predicted the data labels in exp 2 using exp 1 for training data. Note that the feature values and algorithm are the same as used for cross-validation in the previous subsection. In particular, the relationship between crowd density and feature values of the passer in exp 2 is predicted using data for exp 1. 

Table \ref{accuracy_test} shows the results of the prediction. The sensitivity of label \(H\) remained as high as it did in cross-validation using data from exp 1. Therefore, it can be said that our method for estimating crowd density is little influenced by differences between individual passers when detecting high-density situations. Note that the arbitrary parameters in our method are the feature values, \(m\) and \(k\). We consider the detection performance in high-density situations to be unattributable to the hyperparameters \(m\) and \(k\) because these hyperparameters are not related to crowd density; thus, our method is suitable for detecting high density situation due to the choice of feature values.

On the other hand, compared with the cross-validation in the previous subsection, the accuracy was low overall because the percentage of data with label \(H\) was low in the test data (exp 2). We expect that our method will be adapted to detection of high-density situations since the sensitivity of label \(H\) was \(100\%\) except when non-sequential velocity was used; thus, higher percentages of test data with label \(H\) may lead the higher accuracy. This consideration is supported by the fact that the accuracy of the prediction using non-sequential velocity was high, making it suitable for detecting low-density situations.

Although the precision of prediction using wavelet coefficients was worse than that using cross-validation (Sec. 5.3.2), the wavelet-coefficient performance is as good as that using raw velocity. Therefore, we reconfirmed that the wavelet coefficient was effective in our method.

Summarizing Sec. 5.3, we can detect high-density situations 100\% of the time using most feature values. Based on the fundamental diagram, superior performance is natural in the case of using raw velocity. The performance in the sequential-velocity case was better than that using raw velocity due to the number of dimensions. On the other hand, the accuracies using the wavelet coefficient and maximum angular velocity were equal to or higher than that using raw velocity, meaning that angular velocity can be used to estimate crowd density if we properly extract feature values. This implies that there is a relationship between angular velocity and crowd density, as well as between velocity and crowd density.

\begin{table}[htb]
  \centering
  \caption{Validation results using exp 2}
\scalebox{0.9}{
\begin{tabular}{p{0.01\linewidth}p{0.29\linewidth}p{0.11\linewidth}p{0.11\linewidth}p{0.11\linewidth}p{0.11\linewidth}p{0.11\linewidth}}
 \({\bf x}_i\) &     Feature value & Accuracy [\%]  & Sensitivity of \(L\) [\%] & Sensitivity of \(H\) [\%]  & Precision of \(L\) [\%] & Precision of \(H\) [\%] \\ \hline \hline
 \({\bf x}_1\) &    raw velocity & 66.6 & 50.0 & 100.0 & 100.0 & 50.0\\ 
 \({\bf x}_2\) &    raw angular velocity & 33.3 & 0 & 100.0 & - & 33.3 \\  
 \({\bf x}_3\) &    maximum angular velocity & 77.8 & 66.7 & 100.0 & 100.0 & 60.0 \\ 
 \({\bf x}_4\) &    wavelet coefficient & 66.6 & 50.0 & 100.0 & 100.0 & 50.0 \\
 \({\bf x}_5\) &    sequential velocity & 72.2 & 58.3 & 100.0 & 100.0 & 54.5  \\ 
 \({\bf x}_6\) &    non-sequential velocity & 82.4 & 83.3 & 80.0 & 90.9 & 66.7  \\
\end{tabular}
\label{accuracy_test}
}
\end{table}


\section{Local density}
\subsection{Regression}
As with global density, we conducted two validations, namely cross-validation using the data in exp 1 and another validation regarding exp-2 data as test data. Leave-one-out cross-validation (\(m=240\)) is employed and we used \(k\)-NN algorithms as regression problems, where \(k=51\). Note that weighting was introduced to the \(k\)-NN algorithm. 

Predicted labels are the local densities at each time step. Therefore, we estimated the performance of our algorithm by comparing the true and predicted labels at each time step. Note that we have omitted the maximum angular velocity \({\bf x}_3\) and the non-sequential velocity \({\bf x}_6\) in this section, because the two feature values do not correspond to a specific moment; thus, we consider them unsuitable for predicting local density. 

\subsection{TP graph}
In order to visualize the performance of regression predictions, we introduced a TP graph whose horizontal axis represents predicted labels and whose vertical axis represents true labels. For example, in the case where the predicted label is \(2\ [{\rm m}^{-2}]\) and the true label is \(3\ [{\rm m}^{-2}]\) at a certain time step, the point \((2,\ 3)\) is plotted on the graph. In the same manner, the prediction results for all the test data were plotted on the TP graph. The closer to a straight line with slope 1 and intercept 0 the distribution of these points is, the more accurate the estimation model is.

We calculated the root-mean-square error (RMSE), which is an index estimating the accuracy of predictions quantitatively and helping to compare feature values with each other.

\begin{equation}
{\rm RMSE} = \sqrt{\frac{\sum (y_{\rm true} - y_{\rm pred})^2}{N_{\rm all}}}
\end{equation}
\noindent
The RMSE is an indicator of errors between predicted and true labels, and always has a positive value.

\subsection{Results and Discussion}
\subsubsection{Plots of predicted density and true density}
\begin{figure}[h]
	\begin{center}
		\includegraphics[width=\linewidth]{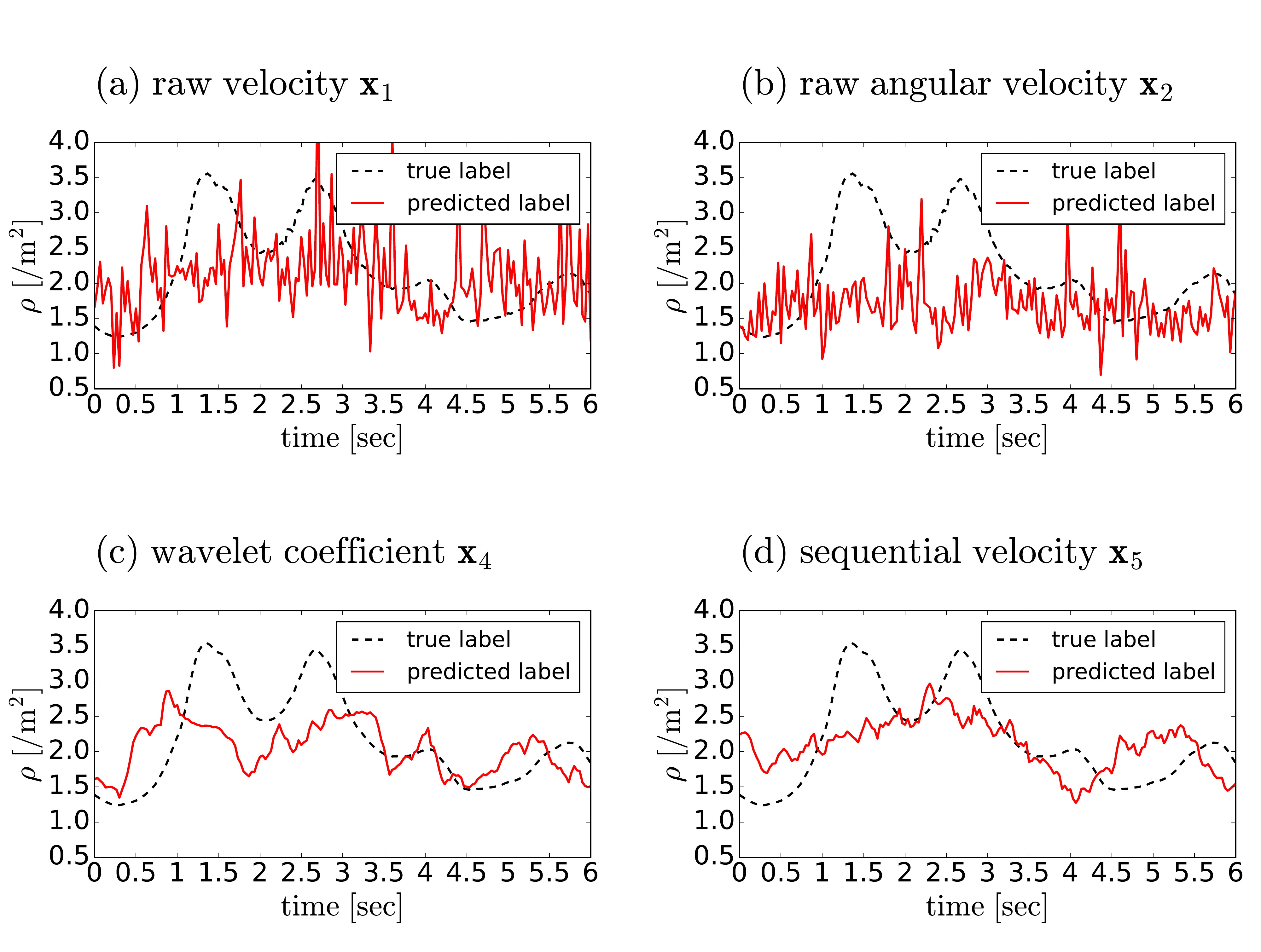}
	\end{center}
\caption{The results of cross-validation using exp-1 data. (a) raw velocity \({\bf x}_1\); (b) raw angular velocity \({\bf x}_2\); (c) wavelet coefficient \({\bf x}_4\); (d) sequential velocity \({\bf x}_5\). The black dashed curve represents the true labels and the red solid curve represents the predicted ones. The number of crowd members was 40.}
	\label{PT_time_series}
\end{figure}

In order to investigate the true and predicted labels, we plotted them as shown in Fig. \ref{PT_time_series}, when the number of crowd members is 40. Fig. \ref{PT_time_series} shows that the predicted labels obeyed a similar tendency to the true labels, except in the case using the raw angular velocity. Although prediction by raw angular velocity fluctuated (Fig. \ref{PT_time_series}(b)), the labels predicted using the wavelet coefficient fluctuated less (Fig. \ref{PT_time_series}(c)). Moreover, there were two peaks in the wavelet-coefficient prediction, corresponding to the peaks shown in the true label time series around 1.4 [sec] and 2.6 [sec]. In Fig. \ref{PT_time_series}(a), although the difference between the predicted and true labels seems as small as that in the Fig. \ref{PT_time_series}(c), the fluctuation is strong. On the other hand, the predicted label in Fig. \ref{PT_time_series}(d) is as smooth as that in Fig. \ref{PT_time_series}(c), but the two peaks are obscure. Although these discussions only concern qualitative characteristics, we discuss quantitative aspects in the next subsection.


\subsubsection{TP graph: exp-1 data}
\begin{figure}[!h]
	\begin{center}
		\includegraphics[width=\linewidth]{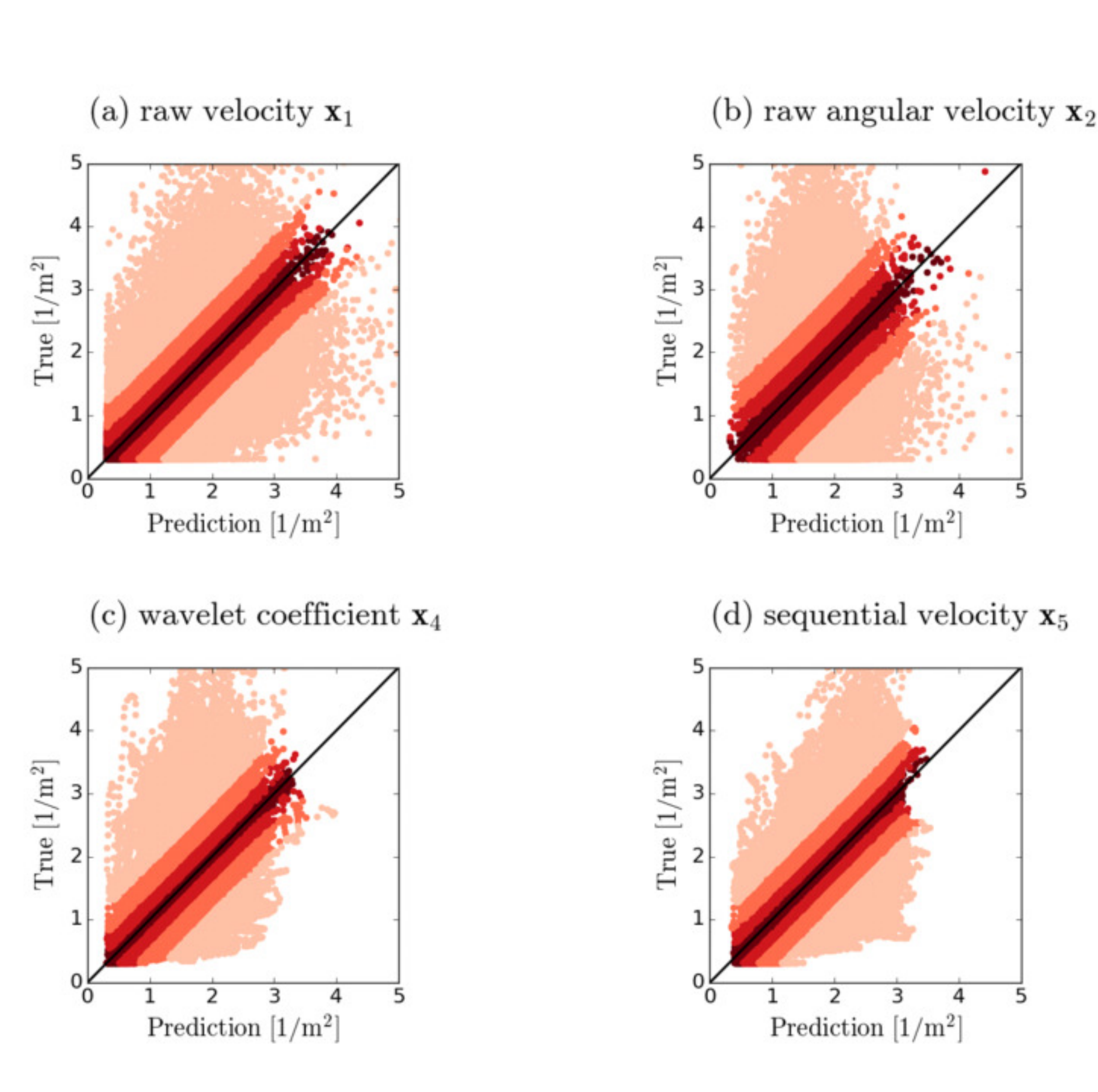}
	\end{center}
\caption{TP graph of each feature value. (a) raw velocity \({\bf x}_1\); (b) raw angular velocity \({\bf x}_2\); (c) wavelet coefficient \({\bf x}_4\); (d) sequential velocity \({\bf x}_5\).}
	\label{PT graph}
\end{figure}

Fig. \ref{PT graph} shows the results of each feature value. Note that we colored the points according to the order of closeness to the line with slope 1 and intercept 0, growing darker as the points approach the line. We gave 4 colors to the points, representing those in the top \(25\%\), the top \(25\sim50\%\), the top \(50\sim75\%\) and the top \(75\sim100\%\), respectively. We focus on the border lines between these areas. The slopes of these lines are almost \(1\), meaning that \(75\%\) of the errors between predicted and true labels falls within a certain range, regardless of the size of the true labels. Therefore, it can be said that the error between the predicted and true labels is independent of the true label. 

We focused on the points in the top \(75\sim100\%\), which were placed in the lower-right and upper-left areas in Fig. \ref{PT graph}. The points in the lower-right area show where the predicted labels far exceeded the true labels. In case using the wavelet coefficient, there were less points than those using other feature values. On the other hand, the points in the upper-left area were observed in every feature value. Therefore, it is difficult to deal with the datasets in this area using our method, while the ratio of datasets in the area was not large; thus, it did not strongly affect the prediction accuracy.

\begin{table}[htb]
  \centering
  \caption{Cross-validation results using exp 1.}
  \begin{tabular} {ccc}
    \({\bf x}_i\) & Feature value & RMSE [\({\rm m}^{-2}\)]   \\ \hline \hline
     \({\bf x}_1\) &raw velocity & 0.84  \\ 
     \({\bf x}_2\) &raw angular velocity & 0.98  \\ 
     \({\bf x}_4\) &wavelet coefficient & 0.84  \\ 
     \({\bf x}_5\) &sequential velocity & 0.73 \\ 
\end{tabular}
\label{RMSE}
\end{table}

Table \ref{RMSE} shows the RMSE of each feature value. It can be said that prediction using the wavelet coefficient was more effective than that using all other feature values besides the sequential velocity. This fact implies that the CWT properly extracted the characteristics of the angular-velocity time-series data. In addition, we found that the accuracy using the wavelet coefficient is at the same level as that using the raw velocity, where the relationship between crowd density and the velocity is shown in fundamental diagrams.


On the other hand, we compared the raw and sequential velocities to estimate the influence of different numbers of feature-vector dimensions. Table \ref{RMSE} shows that the RMSE using sequential velocity was less than that using raw velocity. Therefore, we believe that the velocity data should be handled not as instantaneous values, but as sequential data. 

\subsubsection{TP graph: data for exps 1 and 2}
\begin{figure}[H]
	\begin{center}
		\includegraphics[width=\linewidth]{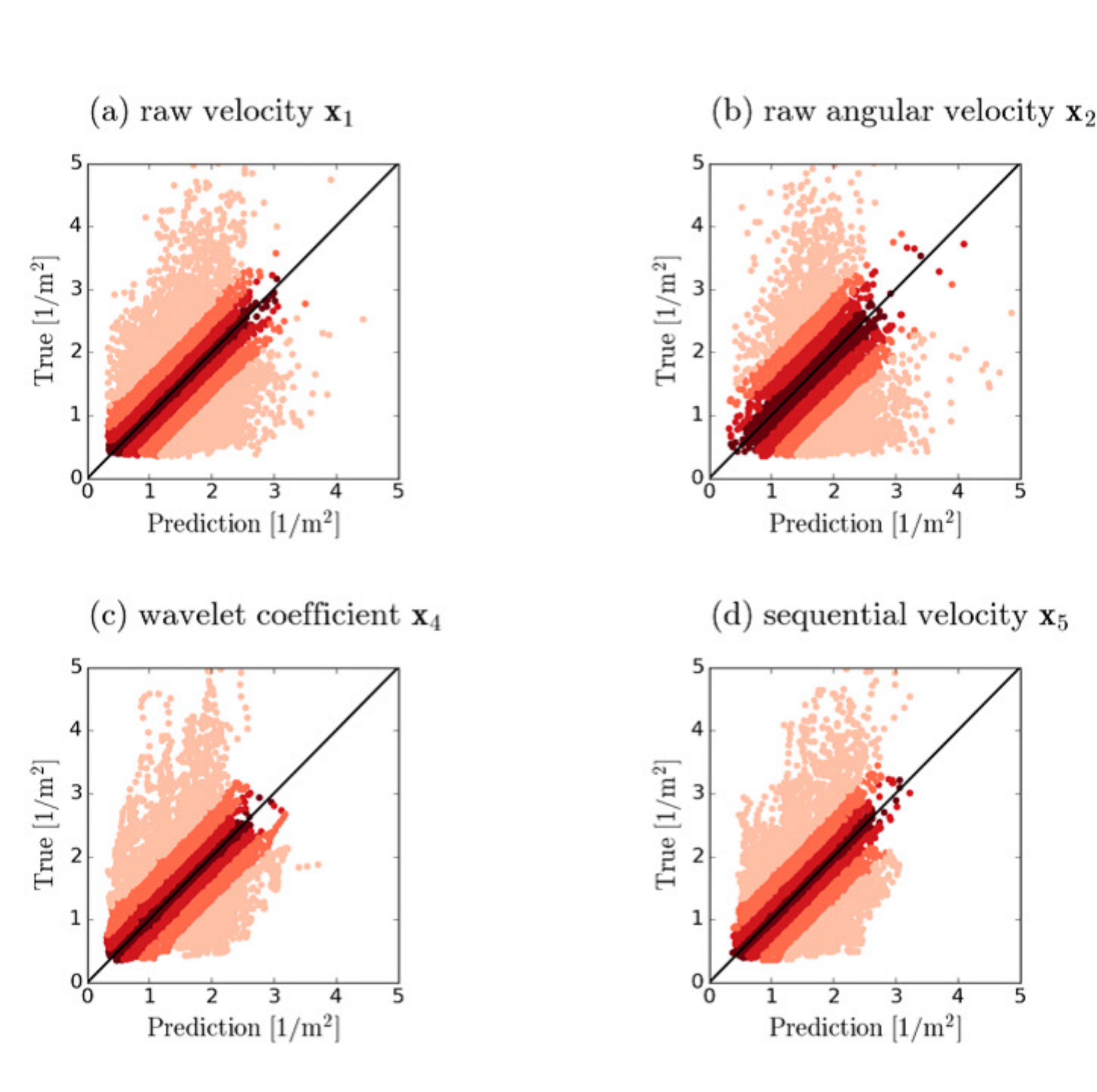}
	\end{center}
\caption{TP graph for each feature value. (a) raw velocity \({\bf x}_1\); (b) raw angular velocity \({\bf x}_2\); (c) wavelet coefficient \({\bf x}_4\); (d) sequential velocity \({\bf x}_5\). Note that the test data were obtained in exp 2.}
	\label{PT graph test}
\end{figure}

In this subsection, we conducted validation using exp-2 data as test data. Fig. \ref{PT graph test} and Table \ref{RMSE test} are the results of this validation. Table \ref{RMSE test} shows that the prediction using the wavelet coefficient was more accurate than that using raw angular velocity. These results are consistent with those of cross-validation in the previous subsection, so it can be said that the CWT is effective in our method. 

On the other hand, the predictions using raw or the sequential velocities are a little more accurate than that using the wavelet coefficient. However, the accuracy difference was too small for us to decide whether the difference is derived from sampling error or other contributing factors. To identify the main factor, additional experiments in which participants are changed from exp 1 to exp 2 are necessary. Moreover, it is easier to measure angular velocity than velocity of pedestrians; thus, we believe that wavelet coefficient \({\bf x}_4\) is useful in application.

The RMSE of each feature value did not increase drastically from that obtained by the cross-validation in the previous subsection. This suggests that our method is little affected by differences in individual passers. This enhances the reliability of our method for estimating crowd density when we apply it to various kinds of pedestrians in the real world.

In our method applying wavelet transform, the accuracy of the prediction is the same as that using raw velocity data. Based on the strong relationship between velocity and crowd density, we considered prediction by the wavelet coefficient to be sufficiently accurate in practice. Conversely, the fact that prediction using angular-velocity data was sufficiently accurate and robust implied a background relationship between angular velocity and crowd density. In the future, further understanding of the angular velocity of pedestrians may allow us to find new aspects of pedestrian dynamics.

\begin{table}[htb]
  \centering
  \caption{Validation results using exp 2}
  \begin{tabular} {ccc}
    \({\bf x}_i\) & Feature value & RMSE [\({\rm m}^{-2}\)]   \\ \hline \hline
    \({\bf x}_1\) &raw velocity & 0.71  \\  
    \({\bf x}_2\) &raw angular velocity & 0.87 \\ 
    \({\bf x}_4\) & wavelet coefficient & 0.75  \\ 
    \({\bf x}_5\) & sequential velocity & 0.71 \\ 
\end{tabular}
\label{RMSE test}
\end{table}

\section{Conclusion}
It is important to measure crowd density to calculate the LOS or to expand an understanding of pedestrian dynamics. Conventionally, pedestrians have been recorded via video camera and then counted. However, this measuring method 
is time-consuming because this type of data is not easy to handle automatically. Moreover, recordable area is usually limited by the design of facilities or laws relating to the right to privacy. Therefore, we focused on data obtained by tablet sensors with which pedestrians are equipped. We developed a new method for estimating crowd density using time-series sensor data; this method offers advantages in terms of cost.

Firstly, we conducted controlled experiments in a square area defined by belt partitions (exp 1). The participants (crowd members) walked randomly in the experimental area as pedestrians in a public space. Another participant (the passer), equipped with a tablet on their torso, walked through the crowd. We varied the number of crowd members and reproduced various crowd densities. The simple experiments gave us the time-series angular-velocity data for the passer when passing through the crowded area. 

Next, we applied a CWT to the obtained data to extract feature values. Using the CWT, we can collect information about the locations and frequency of the signal. Therefore, we could intuitively understand dynamic changes in the angular-velocity data.

Then, we divided the data into two groups, training data and test data, and predicted the crowd density by applying the \(k\)-nearest-neighbor (\(k\)-NN) machine learning algorithm to the extracted feature values (estimation 1). We considered two definitions of density: global density and local density. We regarded the prediction of the global and the local densities as classification and regression, respectively. Using the global density, we can decide whether or not the crowd density is generally high. On the other hand, using local density, we can focus on the relationship between the movement of a passer and the degree of congestion felt by the passer in detail. Our density-predicting method was found to be effective in both cases. In addition, in the case of local density, the estimation error using the wavelet coefficient of angular velocity was equivalent to the error using the raw velocity. This fact means that angular velocity is related to the crowd density, just as velocity is. 

In order to investigate whether differences in individual pedestrians influence the above results, we conducted similar experiments in which different students participated on different days (exp 2). Then, we predicted the crowd density of trials in exp 2 based on data in exp 1 using a similar algorithm (estimation 2). Consequently, we found that the sensitivities of the high-density situations were the same for estimations 1 and 2. Therefore, our method for estimating crowd density is little influenced by differences between individual passers in detecting high-density situations, making it advantageous for social applications. 

Note that we consider that our method can be used even when the posture of the tablets is not the same as in our case. From the accelerometer data in the tablets, we track the acceleration of \(x\)-, \(y\)-, and \(z\)- axes, and then calculate the direction of the gravitational acceleration. By coordinate transformation, we can apply our method to the various situations. However, the accuracy of our method is uncertain when the tablet is not fixed to bodies, e.g., for a case wherein a tablet is in a handbag. Therefore, we should try additional cases by varying the method of  carrying tablets to diversify the understanding of our method.

Our study was a basic investigation of whether the angular velocity and crowd density are related; hence, the experimental setting was simple, wherein the crowd members walked randomly. Because we often observe complex combination of random and unidirectional flow in real crowd, we consider that a randomly walking crowd is an extreme pattern, wherein we observed large angular velocity. Therefore, the angular velocities observed in a real crowd may be smaller than those obtained in our experiments. To investigate the effect of the crowd's walking pattern on the relation between angular velocity and crowd density accurately, we should conduct similar experiments in other settings.

We expect that the relationships between angular velocity and crowd density will become dearer, and the possibility of application of our method increases through the improvements to this study described in the previous paragraphs. Such improvements may not only lead to the development of more accurate density-estimation techniques, but also discover unnoticed aspects of pedestrian dynamics. 

\section*{Acknowledgements}
This work was supported by JST-Mirai Program Grant Number JPMJMI17D4, Japan and JSPS KAKENHI Grant Numbers JP25287026 and JP15K17583.

  \break
\appendix
\appendixpage

\section{Comparison between the Fourier transform and the CWT.}

\begin{figure}[H]
	\begin{center}
		\includegraphics[width=\linewidth]{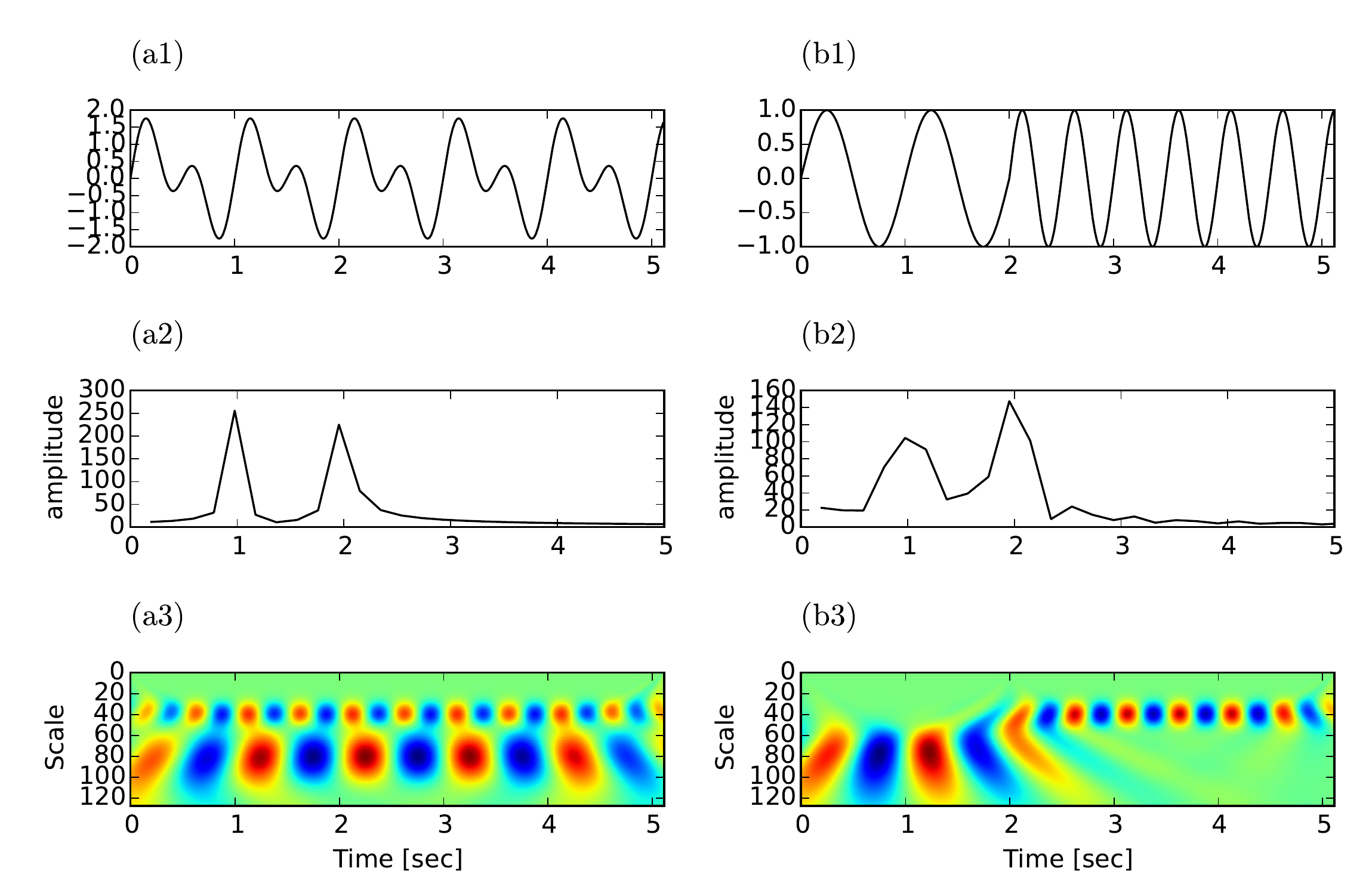}
	\end{center}
\caption{(a1) A 1-Hz and 2-Hz composite wave. (b1) Signal whose frequency is 1 Hz until 2 s and 2 Hz from 2 s. (a2) (b2) Spectra of the two signals, (a1) and (b1), via Fourier transform. (a3) (b3) Wavelet scalograms of the two signals.}
	\label{F_W_comparison}
\end{figure}

We analyzed two signals by using both Fourier transform and CWT, which are important tools for frequency analysis, to prove the advantage of using CWT (Fig. \ref{F_W_comparison}). By applying Fourier transform to the signal in Fig. \ref{F_W_comparison} (a1), we obtain Fig. \ref{F_W_comparison} (a2), whose \(x\)-axis represents frequency and \(y\)-axis represents the strength of each frequency. We can see that Fig. \ref{F_W_comparison} (a1) is mainly comprises of 1-Hz and 2-Hz waves, but it does not include time-based information. Therefore, both Figs. \ref{F_W_comparison} (a2) and (b2) are similar. We can see two peaks with frequencies 1 Hz and 2 Hz. However, CWT also considers the relation between time and frequency, as shown in Eq. \ref{CWT_def}. Therefore, Figs. \ref{F_W_comparison} (a3) and (b3) have distinguishable characteristics. In Fig. \ref{F_W_comparison} (a3), the pattern remains unchanged, whereas it changes at 2 s in Fig. \ref{F_W_comparison} (b3). Thus, we can obtain the frequency changes at 2 s in Fig. \ref{F_W_comparison} (b1) from Fig. \ref{F_W_comparison} (b3).

In our study, the time-based information was necessary because we investigated the local density; thus, we used the CWT.
\section{Influence of changing parameters upon the performance.}
We looked at the influence of changing parameters upon the performance of prediction using wavelet coefficient \({\bf x}_4\). There are two parameters which can be adjusted in this study; the first is \(k\), which represents the number of nearest neighbors in the \(k\)-NN algorithm. The second is \(l\), which corresponds to the dimension of wavelet coefficient \({\bf x}_4\), sequential velocity \({\bf x}_5\), and non-sequential velocity \({\bf x}_6\).

\subsection{Global density}
\begin{figure}[H]
	\begin{center}
		\includegraphics[width=\linewidth]{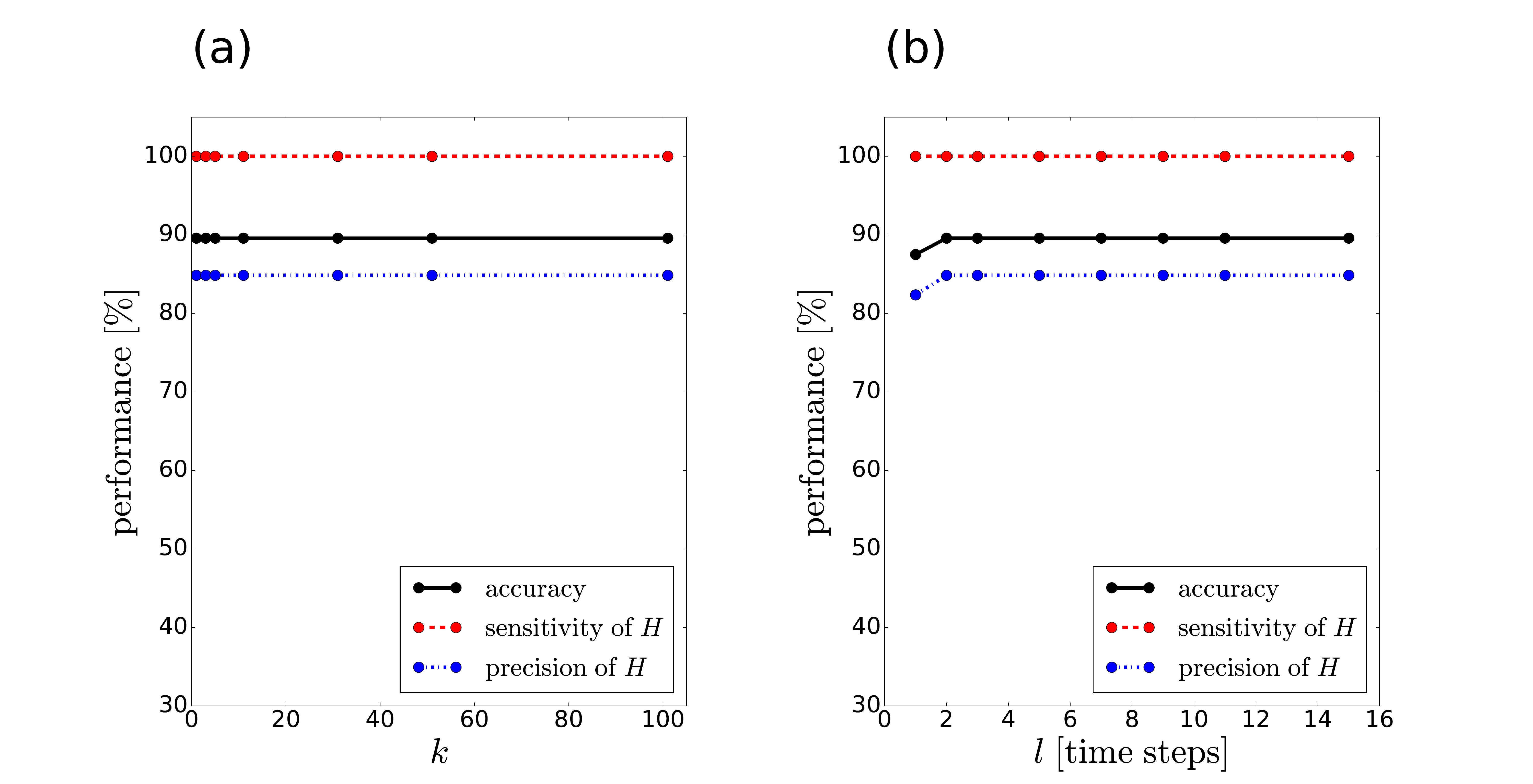}
	\end{center}
\caption{Cross-validation results using exp 1. We plotted the accuracy, sensitivity and precision under variation of (a) \(k\); (b) \(l\).}
	\label{GD_performance}
\end{figure}

Fig. \ref{GD_performance} shows that the performance varies little when changing parameters. In Fig. \ref{GD_performance}(b), the performance increases slightly when varying \(l\) from 1 to 2. However, the variation range is small and this seems to be a natural result because a vector with larger \(l\) includes more information. Note that the number of the training data points was 239 in cross-validation.

\begin{figure}[H]
	\begin{center}
		\includegraphics[width=\linewidth]{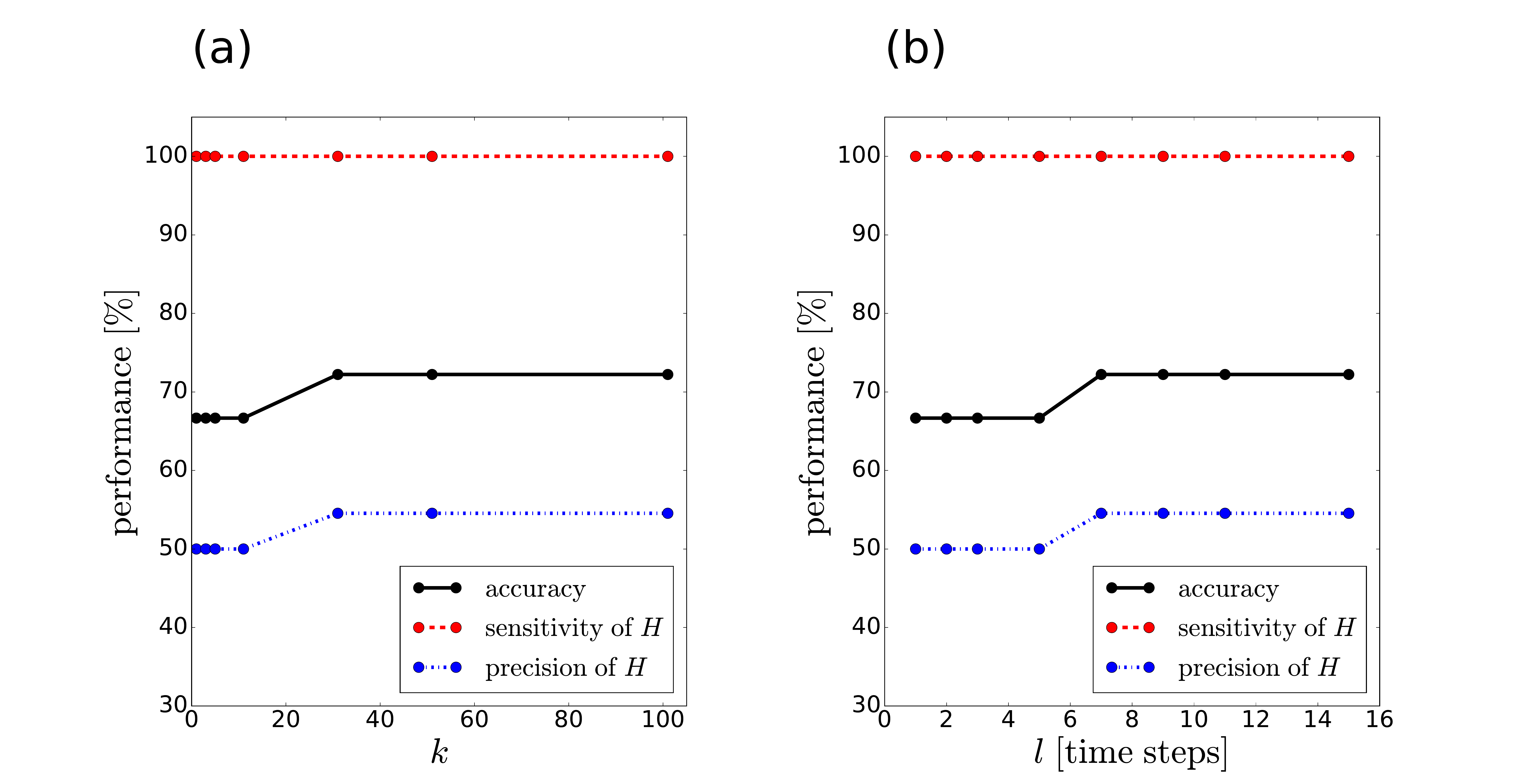}
	\end{center}
\caption{Prediction results in the case of using exp 1 and exp 2. We plotted the accuracy, sensitivity and precision by varying (a) \(k\); (b) \(l\).}
\label{GD_performance_test}
\end{figure}

On the other hand, Fig. \ref{GD_performance_test} shows that the accuracy and precision are smaller than those of Fig. \ref{GD_performance} as mentioned in the main text. However, the sensitivity is 100\% for any parameter values; thus, our method is not influenced by parameters in detecting high-density situations.

Therefore, we considered it reasonable to use \(k=51\) and \(l=7\) when analyzing global density.



\subsection{Local density}
\begin{figure}[H]
	\begin{center}
		\includegraphics[width=\linewidth]{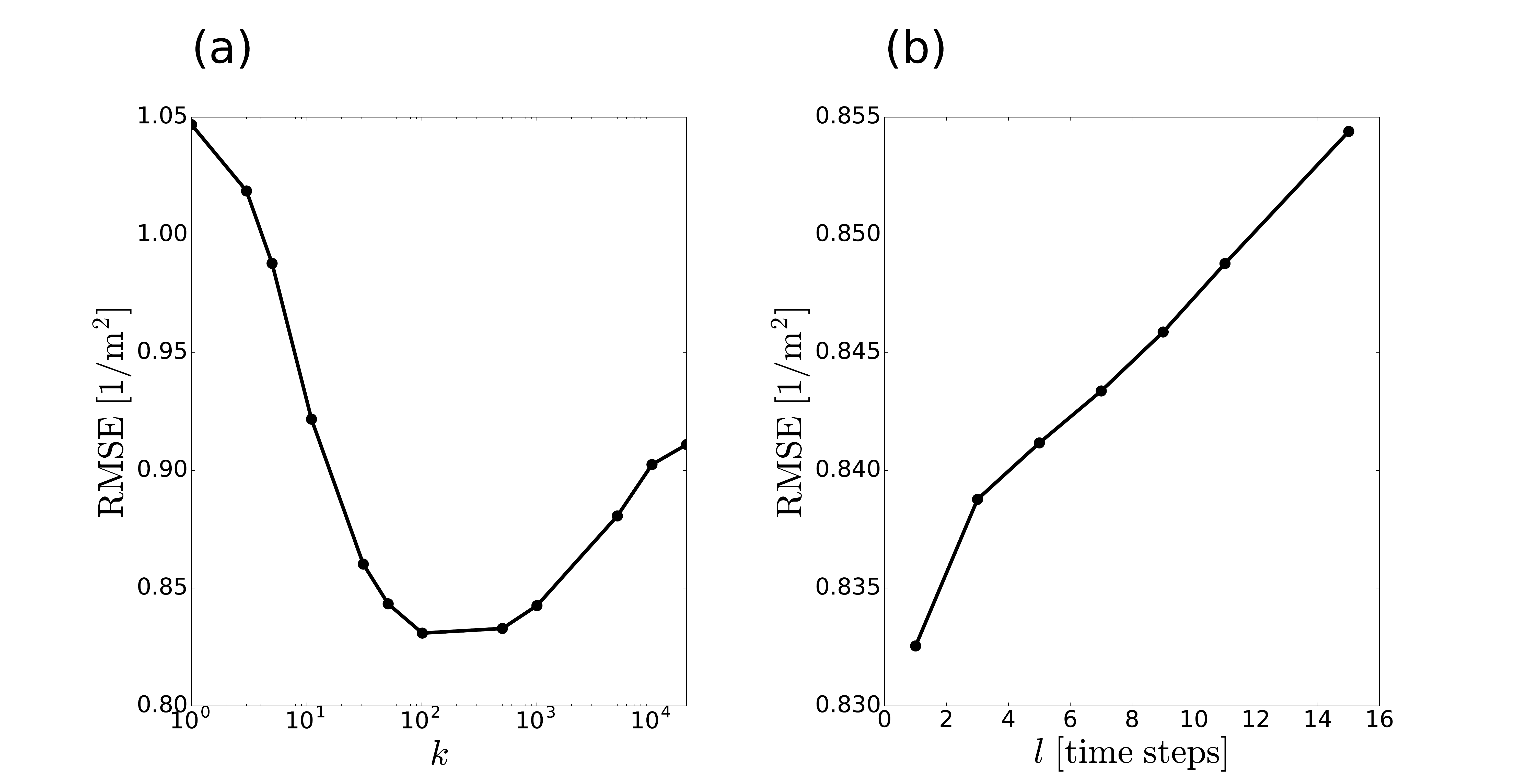}
	\end{center}
\caption{Cross-validation results in the case using exp 1. We plotted the RMSE by varying (a) \(k\); (b) \(l\).}
	\label{LD_performance}
\end{figure}

Fig. \ref{LD_performance}(a) shows that larger \(k\) leads to smaller error in \(k<100\), but the larger error in \(k>1,000\). Fig. \ref{LD_performance}(b) shows that larger \(l\) leads the larger error. However, this variation range is much smaller than that in the case of changing \(k\). Therefore, the choice of \(l\) is not essential to the performance. Note that the number of training data points is about 38000 in the case using exps 1 and 2.

\begin{figure}[H]
	\begin{center}
		\includegraphics[width=\linewidth]{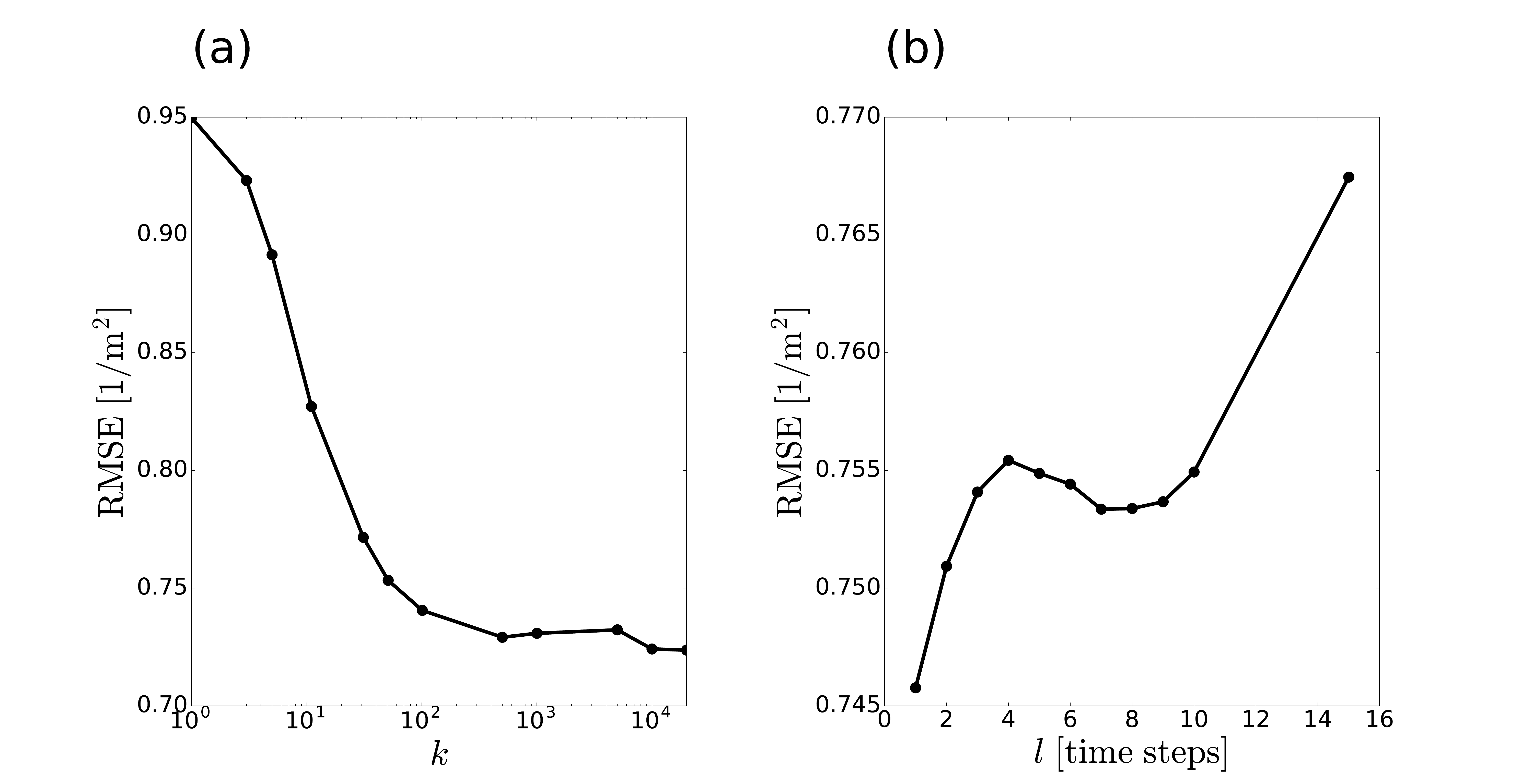}
	\end{center}
\caption{Prediction results using exps 1 and 2. We plotted the RMSE by varying (a) \(k\); (b) \(l\).}
	\label{LD_performance_test}
\end{figure}

On the other hand, Fig. \ref{LD_performance_test}(a) shows that the larger \(k\) leads to smaller error for \(k<100\), as in Fig. \ref{LD_performance}(a). In the case of Fig. \ref{LD_performance_test}(a), there is not obvious minimum value, but the error does not greatly change in \(k>100\). In Fig. \ref{LD_performance_test}(b), there is a maximum and minimum around \(k=4\) and \(k=8\), respectively. However, the variation range is much smaller than that in the case of changing \(k\).

Therefore, we decided to use \(k=51\) and \(l=7\) in analyzing the local density, where the parameter values conformed to those in the global-density analysis.



\end{document}